\documentclass[final,3p,times]{elsarticle}

\usepackage{amssymb}
\usepackage{longtable}
\usepackage{mathtools}
\usepackage{url}
\usepackage{comment}
 \usepackage{amsthm}
\usepackage{amsmath}
\usepackage{multirow}
\usepackage{array,longtable}
\usepackage[dvipsnames]{xcolor}
\usepackage{float}
\usepackage{framed} 
\usepackage[superscript]{cite}

\usepackage{multicol} 

\usepackage{nomencl} 

\makenomenclature

\renewcommand*\nompreamble{\begin{multicols}{2}}

\renewcommand*\nompostamble{\end{multicols}}

\usepackage{etoolbox}
\renewcommand\nomgroup[1]{%
  \item[\bfseries
  \ifstrequal{#1}{S}{Sets and Indices}{%
  \ifstrequal{#1}{P}{Parameters}{%
  \ifstrequal{#1}{V}{Variables}{}}}%
]}

\journal{Nature}

\begin{document}

\begin{frontmatter}



\title{Granular Compensation, Information, and Carbon Pricing Promote DER Deployment}


\author[Tandon]{Hafiz Anwar Ullah Khan \corref{cor1}}
\ead{anwar.khan@nyu.edu}
\author[Law]{Bur\c{c}in \"{U}nel}
\ead{burcin.unel@nyu.edu}
\author[JHU]{Yury Dvorkin}
\ead{ydvorki1@jhu.edu}
\cortext[cor1]{Corresponding Author}

\address[Tandon]{Department of Electrical and Computer Engineering, NYU Tandon School of Engineering, New York University, NY, USA}
\address[Law]{Institute for Policy Integrity, NYU School of Law, New York University, NY, USA }
\address[JHU]{Ralph O'Connor Sustainable Energy Institute, Johns Hopkins University, Baltimore, MD, USA}

\begin{abstract}
The socially efficient deployment  of Distributed Energy Resources (DERs), e.g., rooftop solar, depends on the underlying retail electricity policies. Current debates on DER policies, including Net Energy Metering (NEM) variants, center around developing value-reflective compensation policies that can expedite DER deployment while preventing potential cost shifts between DER adopters and non-adopters. However, these debates mostly ignore the temporally- and spatially- granular value of DERs,  market failures (e.g., information asymmetry among DER stakeholders) and externalities (e.g., carbon-dioxide emissions). In this paper, we develop a game-theoretic approach with information asymmetry to examine efficiency implications of adopting granular DER compensation policies, e.g., value stacks and distributional locational marginal price, instead of NEM  with flat retail rates. We show that granular compensation policies 
result in more efficient market outcomes than under NEM, even in the presence of information asymmetry, thus avoiding the need for interventions. 
Combined with granular DER compensation, carbon pricing  provides the most accurate price signal to DER investors/aggregators, and leads to the highest social welfare. 

\end{abstract}



\end{frontmatter}


\section{Introduction}

Integration of Distributed Energy Resources (DERs) into power systems is critical for the global decarbonization efforts. DER compensation is a key policy for a socially efficient adoption and operation of DERs.
Net Energy Metering (NEM), whereby the energy exported to the grid is compensated at the retail electricity tariff, is the dominant DER compensation policy \cite{DSIRE}. 
However, existing electricity tariffs are still often temporally and spatially invariant, potentially resulting in socially inefficient economic outcomes, e.g., cost shifts from DER-owners to non-DER-owners \cite{nature_OShaughnessy}, excessive incentives to install and operate DERs \cite{Willdan}, and insufficient revenue streams for  power utilities \cite{nature_Borenstein}. Therefore, the structure, and social impacts of NEM on stakeholders (power utilities, DER industry, consumer and environmental advocacy groups, etc.) have been under scrutiny in the recent years \cite{Burcin_NEM}.

This scrutiny is motivated by a consensus among stakeholders that the rooftop solar compensation rates need to be revised \cite{PAO_testimony, Sylwia}. For example, California is revising its existing rooftop solar compensation mechanism, NEM 2.0, with the ‘net billing’ scheme \cite{net_billing}. However, proposed reforms such as levying an upfront fee for installing solar systems, monthly grid participation charge, ‘glide path’ or retroactive reduction in NEM rates \cite{net_billing} have rekindled the debates among stakeholders about the distributional impacts of such policies. FERC Order 2222 \cite{FERC222} mandating grid operators to allow for DER  participation in the wholesale market led to additional debates about DER compensation. 

Current policy discussions and literature focus primarily on improving the existing NEM structure, but lack nuance in introducing granular compensation policies and accounting for market failures, such as information asymmetry between  DER stakeholders. Bialek \textit{et al.} investigate information barriers to efficient DER roll-out using expert interviews and a stakeholder survey from 13 state-level proceedings on DER compensation  \cite{Sylwia}. The results demonstrate that the effectiveness of DER compensation policies is  affected by information asymmetry in hosting capacity and granular consumer information between the power utilities and DER aggregators \cite{Sylwia}. Hence, addressing information asymmetry in the design of NEM and other granular compensation policies is critical for a socially optimal integration of DERs.

\begin{figure}[!t]
\centering
\includegraphics[scale=0.8]{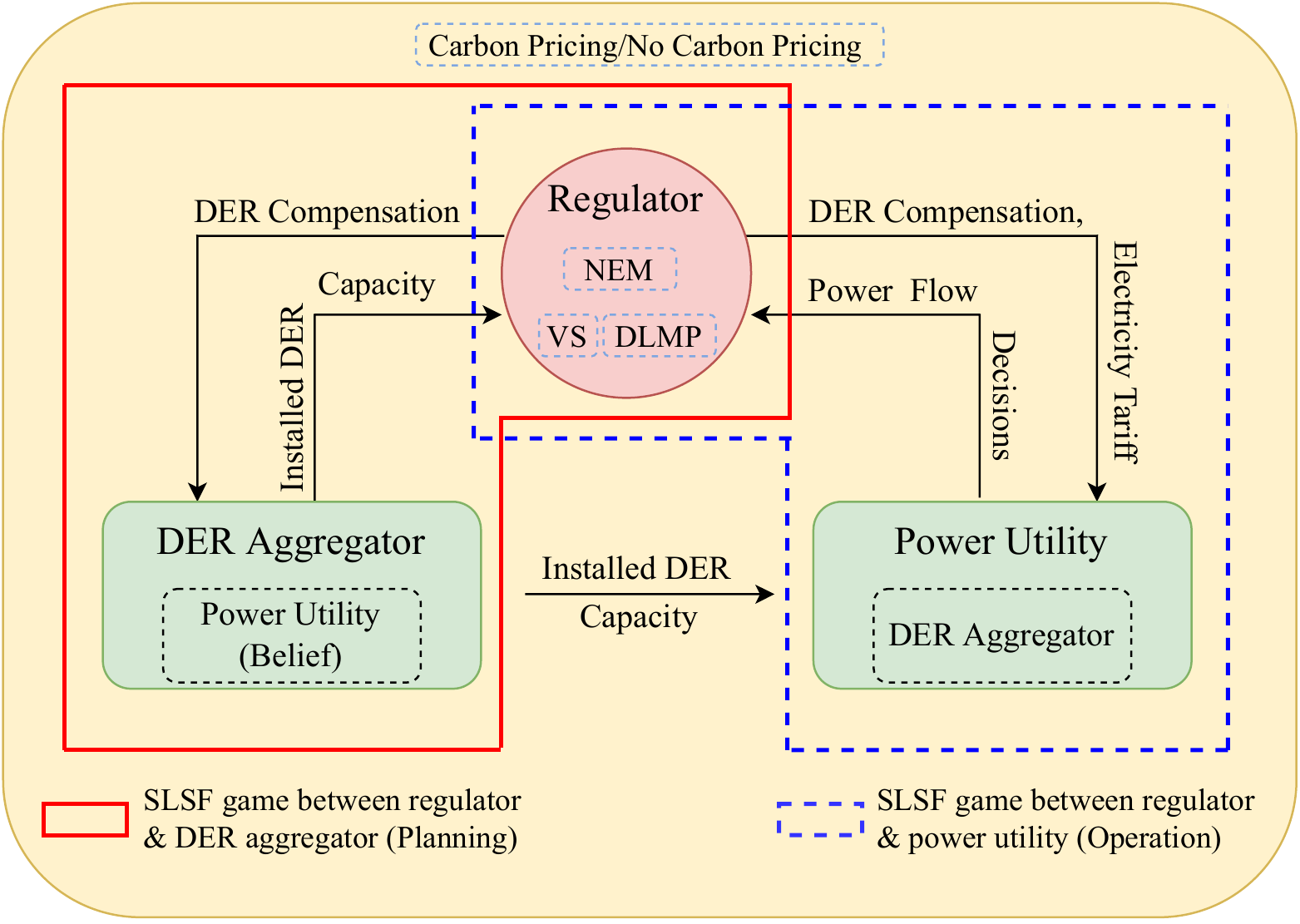}
\caption{\small{A schematic representation of the SLSF games between the regulator and the DER aggregator, and the regulator and the power utility. DER compensation policy, i.e., NEM, VS, or DLMP, is chosen by the regulator. The complete problem is formulated with and without carbon pricing in the system.}}
\label{abstract}
\end{figure}

This paper develops game-theoretical model with information asymmetry to analyze  potential efficiency implications of granular DER compensation policies, such as Value Stack (VS) \cite{NYSERDA} and Distributional Locational Marginal Pricing (DLMP), in comparison to NEM. Our model assumes regulator as the leader and develops two Single Leader Single Follower (SLSF) games with DER aggregator and power utility as the followers in the first and second games, respectively, see Fig.~\ref{abstract}. The regulator maximizes the social welfare, which includes  surpluses of the power utility, consumers, and DER aggregator, and the net monetized value of carbon-dioxide emissions. The power utility and the DER aggregator maximize their profits. The first SLSF game optimizes the installed DER capacity based on the beliefs of the DER aggregator, 
whereas the second SLSF game optimizes the DER compensation and the electricity tariff, subject to the installed DER capacity in the first game.  First, we establish a benchmark case by assuming complete information between the stakeholders and compare NEM, VS, and DLMP for market outcomes (total installed DER capacity, compensation, and social welfare including emissions). Second, motivated by the results in Bialek \textit{et al.}\cite{Sylwia}, we analyze the effects of information asymmetry in hosting capacity and consumer information on the efficiency implications of the three compensation policies. Information asymmetry is modeled as the difference between the actual value and the beliefs of the DER aggregator about system parameters known to the utility. The beliefs are regarded optimistic (pessimistic) if the value of the belief function is greater (lower) than the actual value of the system parameter.  Finally, we introduce carbon pricing, and analyze its effects on the obtained results. 

\begin{figure}[!t]
\centering
\includegraphics[height=3.5in, width=2.8in]{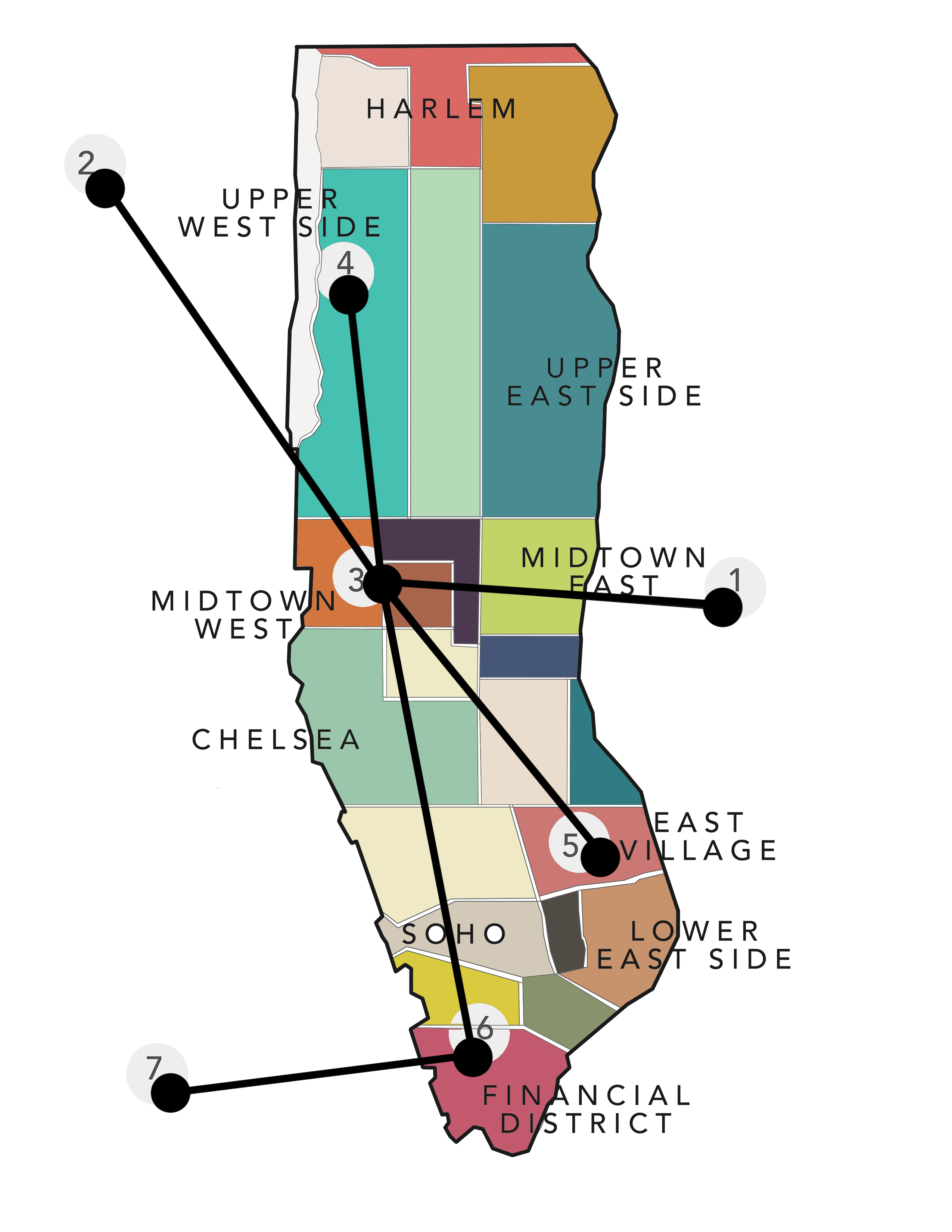}
\caption{\small{A diagram of the 7-zone Manhattan distribution system aggregated in seven nodes (shown in black circles)  with   the transmission system  connected to nodes \# 1, 2 and 7.}}
\label{Manhattan}
\end{figure}

We use a realistic 7-zone power network in Manhattan, NY\cite{SLMF}, as shown in Fig.~\ref{Manhattan}.
We show that the granular DER compensation policies, even in the presence of information asymmetry, outperform NEM in terms of the installed DER capacity and social welfare. 
Moreover, the outcomes are maximized when carbon pricing is introduced along with the granular DER compensation policies. These results provide critical insights for one of the most pertinent DER policy debates.

\section{Effects of DER compensation policies on the market outcomes}
\label{Sec:Results_1}

In this section, we show the effect of 
different compensation schemes on market outcomes under complete information (no information asymmetry). Thus, we compare the widely-employed NEM, the VS adopted in some states \cite{NYSERDA}, and a prospective DLMP-based compensation. The VS is set for a prolonged period of time and  depends on the time and location of the power injected by the DER into the grid, whereas DLMP is calculated by the dual variable of the power balance constraint in the optimal power flow reflecting grid operation conditions close to real-time. To incorporate the impact of carbon pricing on the market outcomes, we define two sub-cases for each compensation policy, whereby the locational marginal price (LMP) in the wholesale market is calculated with and without carbon pricing. 

\begin{figure}[!t]
\centering
\includegraphics[scale=0.28]{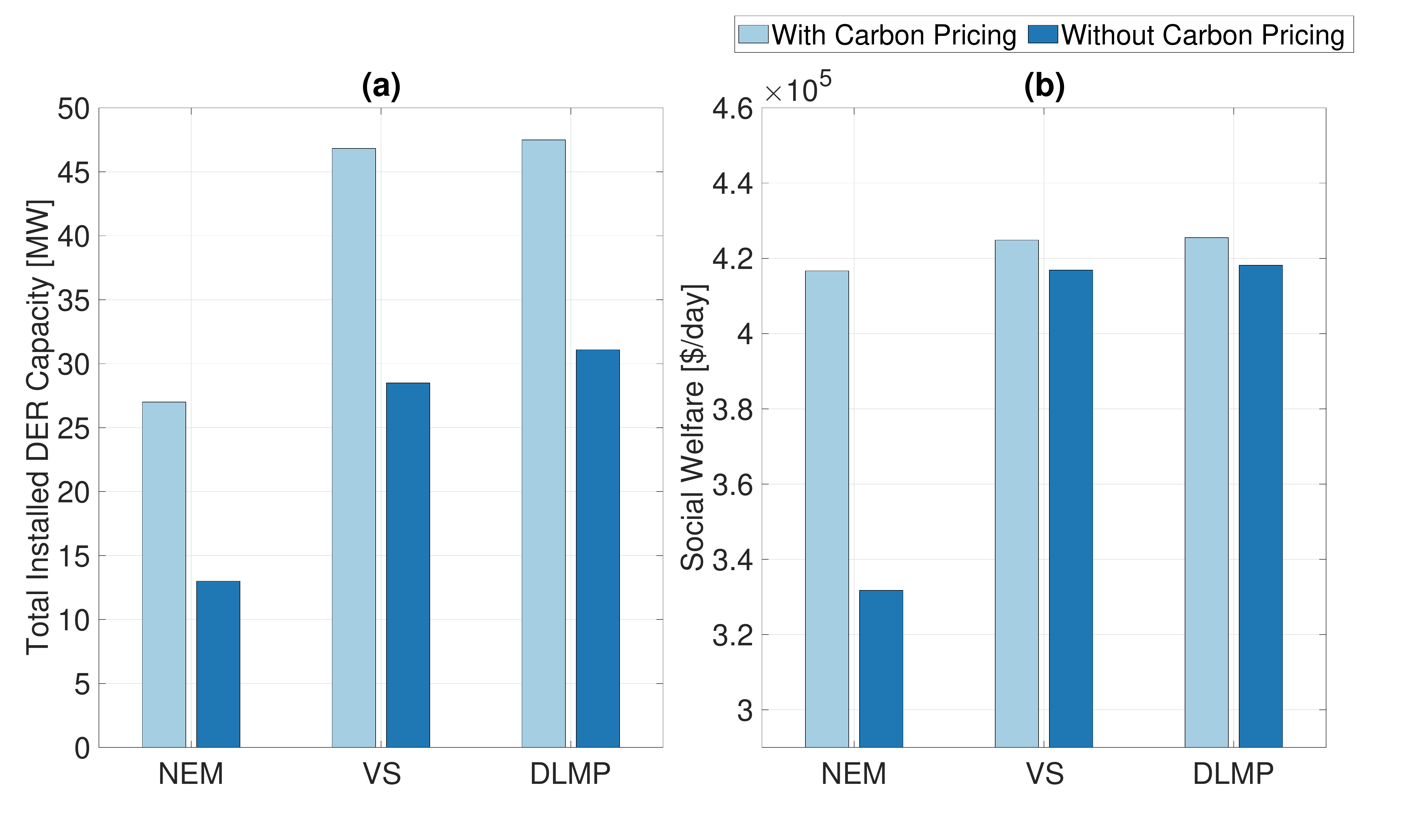}
\caption{\small{Comparison of market outcomes for different DER compensation policies, implemented with and without carbon pricing. (a): Total installed DER capacity, (b): Value of system welfare, calculated using the surpluses of the power utility, consumers, and DER aggregator, and the net monetized value of carbon-dioxide emissions. 
}}
\label{results}
\end{figure}

Figs.~\ref{results}(a) and (b) compare the total installed DER capacity and social welfare under NEM, VS, and DLMP. The DER capacity and social welfare increase under VS and DLMP, as compared to NEM. While differences in the outcomes are minimal between VS and DLMP, their introduction instead of NEM leads to an increase in the installed capacity (up to 54.4\%) and system welfare (up to 20.4\%). Hence, even if  the ideally granular DLMP policy is not implemented, an approximation of the underlying value of DERs, as in VS, can significantly increase the efficiency of the market outcomes. 

\begin{figure}[!t]
\centering
\includegraphics[scale=0.25]{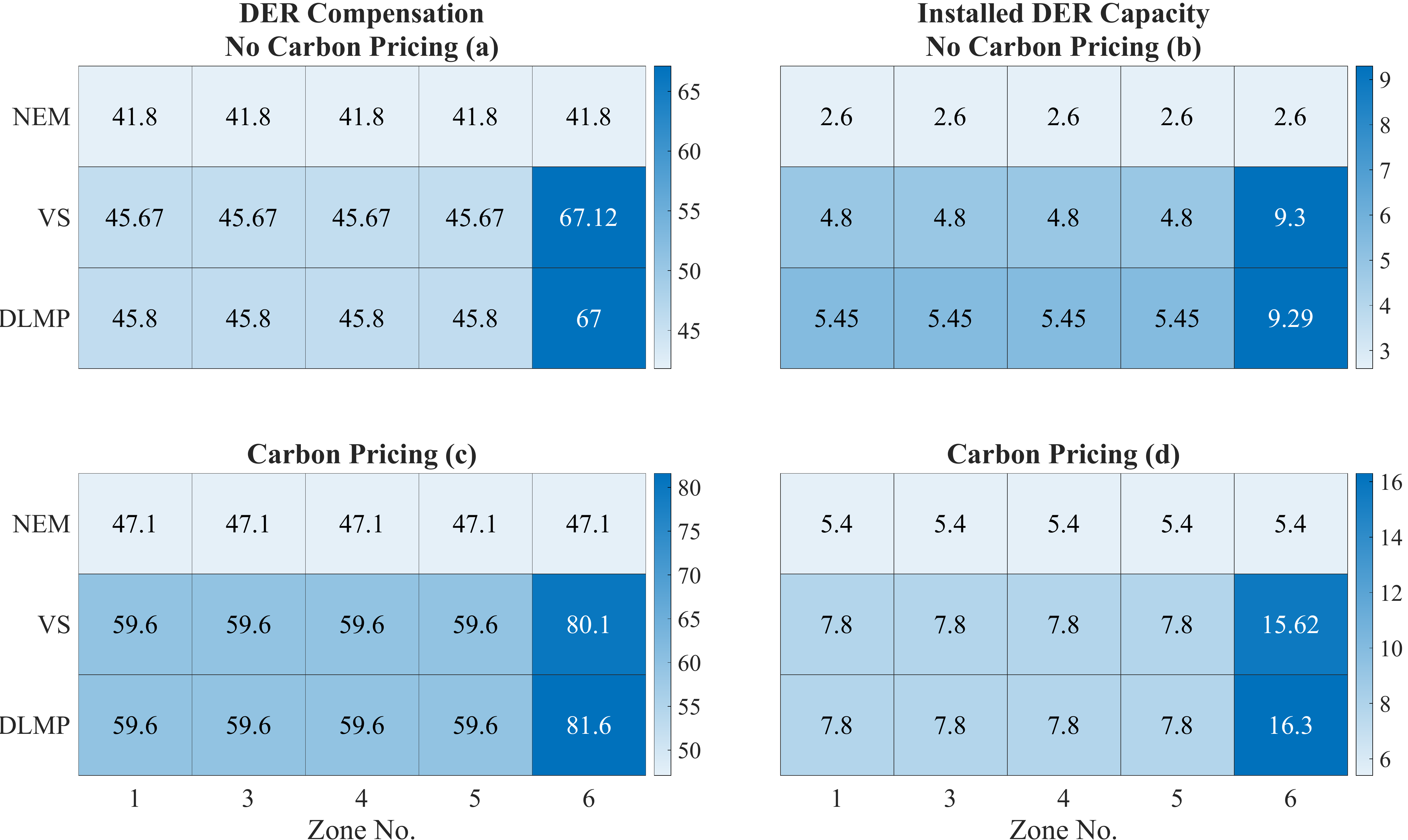}
\caption{\small{Optimal values of DER compensation (\$/MWh) and installed DER capacity (MW) at each zone of Manhattan, for NEM, VS, and DLMP. (a) \& (b): Carbon pricing is not incorporated in the policies, (c) \& (d): Policies include carbon pricing.
}}
\label{Case 1_bus}
\end{figure}

Furthermore, replacing NEM leads to  spatial differences in the values of compensation and the installed DER capacity in  Manhattan, see Fig.~\ref{Case 1_bus}. While for NEM, the DER compensation and capacity at each zone remain the same,  VS and DLMP lead to a higher DER capacity and compensation at zone \# 6, see Figs.~\ref{Case 1_bus}(a) and (b). The higher installed capacity is attributed to the congestion cost of the line connecting zones \# 3 and \# 6.  Owing to a high capacity of the distribution network in Manhattan (i.e., few congestion on other lines), the DER compensation and installed capacity at zones \# 1 -- 5 are the same for both VS and DLMP. The difference in the DER compensation 
and  installed DER capacity 
at zone \# 6 between VS and DLMP, along with a higher social welfare (Fig,~\ref{results}(b)), points to a higher efficiency of DLMP in comparison to VS. For a more congested system, the installed DER capacity along with the DER compensation at each congested zone would likely increase in the case of DLMP, increasing the difference in the welfare between DLMP and VS. 

To quantify the impact of carbon pricing on the market outcomes, we compare the two sub-cases for each compensation policy, see Fig.~\ref{results}. The installed DER capacity and welfare increase under carbon pricing for all compensation policies and the maximum increase in DER capacity and welfare  is observed for NEM, followed by VS and DLMP. 
However, DLMP with carbon pricing outperforms all other policies in terms of the absolute DER capacity and welfare. VS with carbon pricing constitutes the penultimate case, whereas the installed capacity and welfare are relatively low for NEM. The installed DER capacity increases by 72.6\%, and the welfare by 22\%, if the DER compensation policy shifts to carbon pricing-based DLMP as compared to the status quo (NEM without carbon pricing). However, this is an ideal case where both the maximum granularity in compensation and carbon pricing are simultaneously introduced. Even if this drastic policy shift is not possible, replacing NEM with VS (without carbon pricing) would result in 2.6\% more installed capacity 
as compared to introducing carbon pricing under NEM.   
This result underscores the criticality of granular compensation policies for an economically efficient deployment of DERs. The gained efficiency in market outcomes under granular compensation policies  further increases with carbon pricing.

\section{How information asymmetry affects market outcomes under different DER compensation policies}
\label{Sec:Results_2}

To illustrate the impact of information asymmetry between the DER aggregators and the power utilities, 
we define four cases. Case 1 assumes complete information (see Section~\ref{Sec:Results_1}), Case 2 and 3 analyze the impacts of information asymmetry in hosting capacity and consumer information, respectively, and Case 4  considers information asymmetry in both hosting capacity and consumer information. We define two sub-cases for each case to account for the pessimistic (value of belief function $<$ actual value) and optimistic beliefs (value of belief function $>$ actual value) of the DER aggregator. The case with complete information under NEM is our benchmark.

\subsection{Information asymmetry in hosting capacity}
\begin{figure}[!t]
\centering
\includegraphics[scale=0.3]{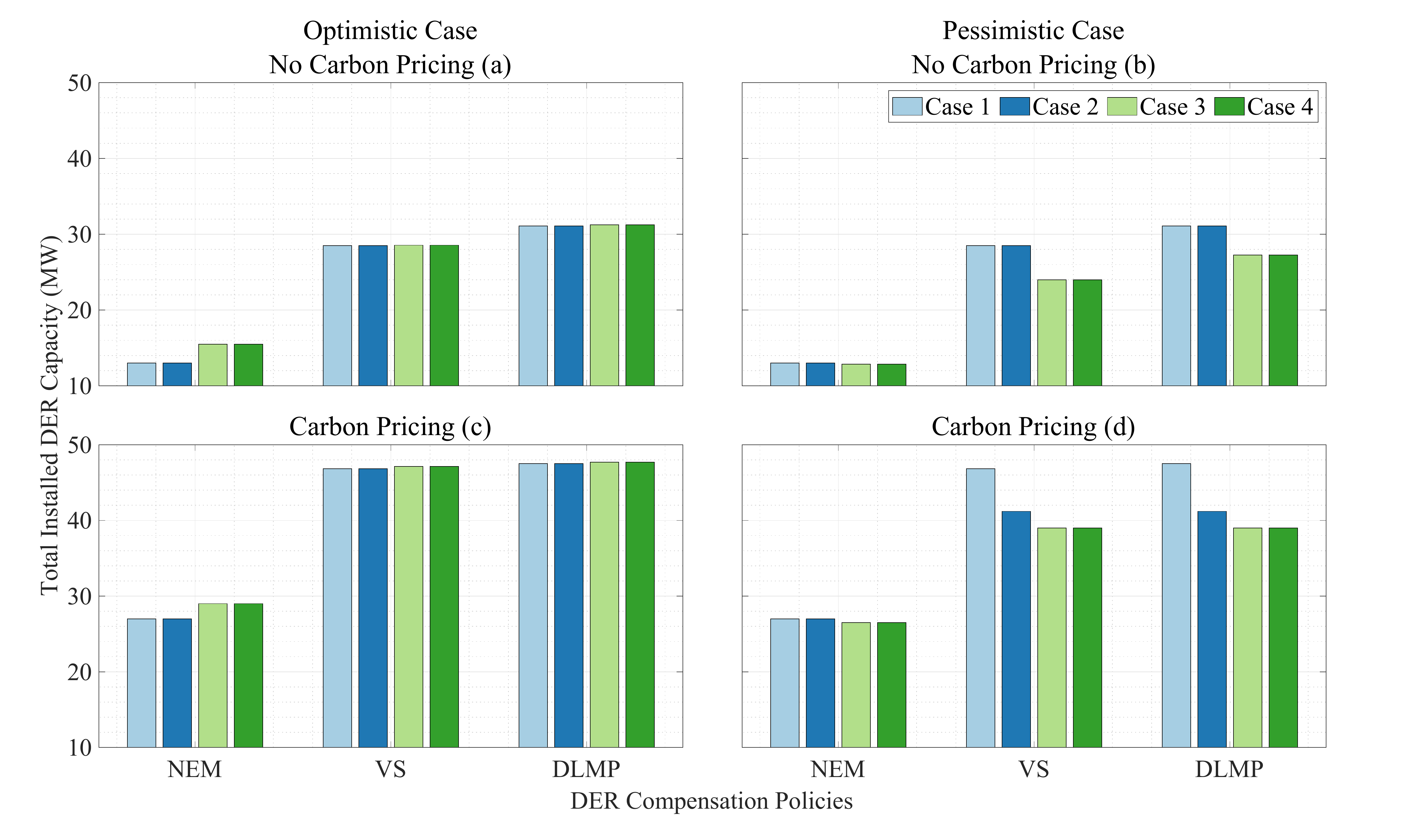}
\caption{\small{Total installed DER capacity in the system incorporating market failures under different DER compensation policies. (a) \& (c): Optimistic beliefs of the DER aggregator under NEM, VS, and DLMP with and without carbon pricing, (b) \& (d): Pessimistic beliefs of the DER aggregator under NEM, VS, and DLMP with and without carbon pricing.
}}
\label{Cap_asymm}
\end{figure}

Under NEM, the optimistic beliefs about hosting capacity (Case 2) do not affect the installed DER capacity, as compared to the complete information case (Case 1), see Fig.~\ref{Cap_asymm}(a). A similar trend is observed under VS and DLMP for optimistic beliefs, see Fig.~\ref{Cap_asymm}(a), and under all compensation policies for the pessimistic beliefs, see Fig.~\ref{Cap_asymm}(b). This is because the installed DER capacity for Cases 1 and 2 under all compensation policies is lower than the hosting capacity. Therefore, information asymmetry in hosting capacity does not impact the market outcomes irrespective of the beliefs of the DER aggregator. Comparing the three compensation policies for Case 2, the installed DER capacity increases by 58.2\% under DLMP, as compared to NEM, for both the optimistic and pessimistic beliefs, see Figs.~\ref{Cap_asymm}(a) and (b). Thus, granular compensation increases the installed DER capacity even in the presence of information asymmetry in hosting capacity. 

\subsection{Information asymmetry in consumer information}
Under NEM, the optimistic beliefs about the consumer information (Case 3) increase the installed DER capacity, as compared to Case 1, see Fig.~\ref{Cap_asymm}(a). The same trend is observed under VS and DLMP, however, the increase is minimal. Since the DER aggregator overestimates the system demand while planning the DER capacity, more capacity is installed in the system. Contrary to the optimistic case, the pessimistic belief about the consumer information (Case 3) decreases the installed DER capacity under all compensation policies, as compared to their respective Case 1, see Fig.~\ref{Cap_asymm}(b). The reduction in the DER capacity is attributed to the underestimated system demand by the DER aggregator. The reduction is less pronounced under NEM, as compared to VS and DLMP, because NEM compensates DERs using the electricity tariff. This restricts large discrepancies in DER compensation, and therefore, in the DER capacity. Hence, information asymmetry in consumer information alters the installed DER capacity depending on the beliefs of the DER aggregator: the optimistic (pessimistic) beliefs increase (decrease) the DER capacity in the system. 

Comparing the three DER compensation policies for Case 3, the installed DER capacity increases under VS and DLMP, as compared to NEM, for both the optimistic and pessimistic beliefs, see Figs.~\ref{Cap_asymm}(a) and (b). The increase in DER capacity is higher as we introduce more granular compensation under pessimistic beliefs, as compared to the optimistic beliefs, compare Figs.~\ref{Cap_asymm}(a) and (b). 
We explain the higher increase under pessimistic beliefs with low value of DER capacity installed under NEM for the pessimistic beliefs in Case 3. Thus, temporal and spatial granularity results in value-reflective DER compensation policies, increasing the installed DER capacity even in the presence of information asymmetry in consumer information. 

\subsection{Information asymmetry in hosting capacity and consumer information}
The installed DER capacity does not change in Case 4 as compared to Case 3, under all the three DER compensation policies, irrespective of the beliefs of the DER aggregator, see Figs.~\ref{Cap_asymm}(a) and (b). Owing to the absence of adequate policy incentives to install and operate DERs, the installed DER capacity in Case 3 is extremely low as compared to the hosting capacity at each zone. Hence, the beliefs about the hosting capacity do not affect the results.

\subsection{Effects of carbon pricing}
Figs.~\ref{Cap_asymm}(c) and (d) show the effects of introducing carbon pricing on the installed DER capacity for Cases 1--4 under the three DER compensation policies. For the optimistic beliefs, the trends under carbon pricing remain similar to the trends under no carbon pricing, compare Figs.~\ref{Cap_asymm}(a) and (c). The optimistic beliefs either do not impact (Case 2) or minimally increase (Case 3) the installed DER capacity as compared to the complete information case under all  three compensation policies.  Carbon pricing increases the installed DER capacity for Cases 1--4 under all compensation policies, as compared to the same cases with no carbon pricing, see Figs.~\ref{Cap_asymm}(a) and (c). 

However, for the pessimistic beliefs, the trends with and without carbon pricing for Cases 1--4 under VS and DLMP are not similar, compare Figs.~\ref{Cap_asymm}(b) and (d).  Pessimistic beliefs about the hosting capacity (Case 2) with carbon pricing reduce the installed DER capacity under VS and DLMP as compared to the respective perfect information cases, contrary to the trend observed in Fig.~\ref{Cap_asymm}(b).  Thus, carbon pricing incentivizes the deployment of DERs, resulting in a higher DER capacity at each zone, as compared to the cases without carbon pricing. The pessimistic beliefs about the hosting capacity curtail the DER deployment, constraining the values to the belief function, resulting in a lower installed DER capacity as compared to Case 1, see Fig.~\ref{Cap_asymm}(d). When there is no carbon pricing, see Fig.~\ref{Cap_asymm}(b), the pessimistic beliefs about the hosting capacity do not affect the results, since the installed DER capacity with perfect information  is already lower than with the pessimistic beliefs.

Finally, the effects of information asymmetry are exacerbated for all compensation policies, when carbon pricing is introduced. For example, for the pessimistic beliefs under VS, the installed DER capacity between Cases 1 and 3 decreases by 16.7\% with carbon pricing, as compared to 15.8\% without carbon pricing, compare Figs.~\ref{Cap_asymm}(b) and (d). 
However, the increase in the installed capacity owing to carbon pricing, 
see Figs.~\ref{Cap_asymm}(a) and (c), and \ref{Cap_asymm}(b) and (d), outweighs the exacerbated effects of information asymmetry.

\begin{figure}[!t]
\centering
\includegraphics[scale=0.35]{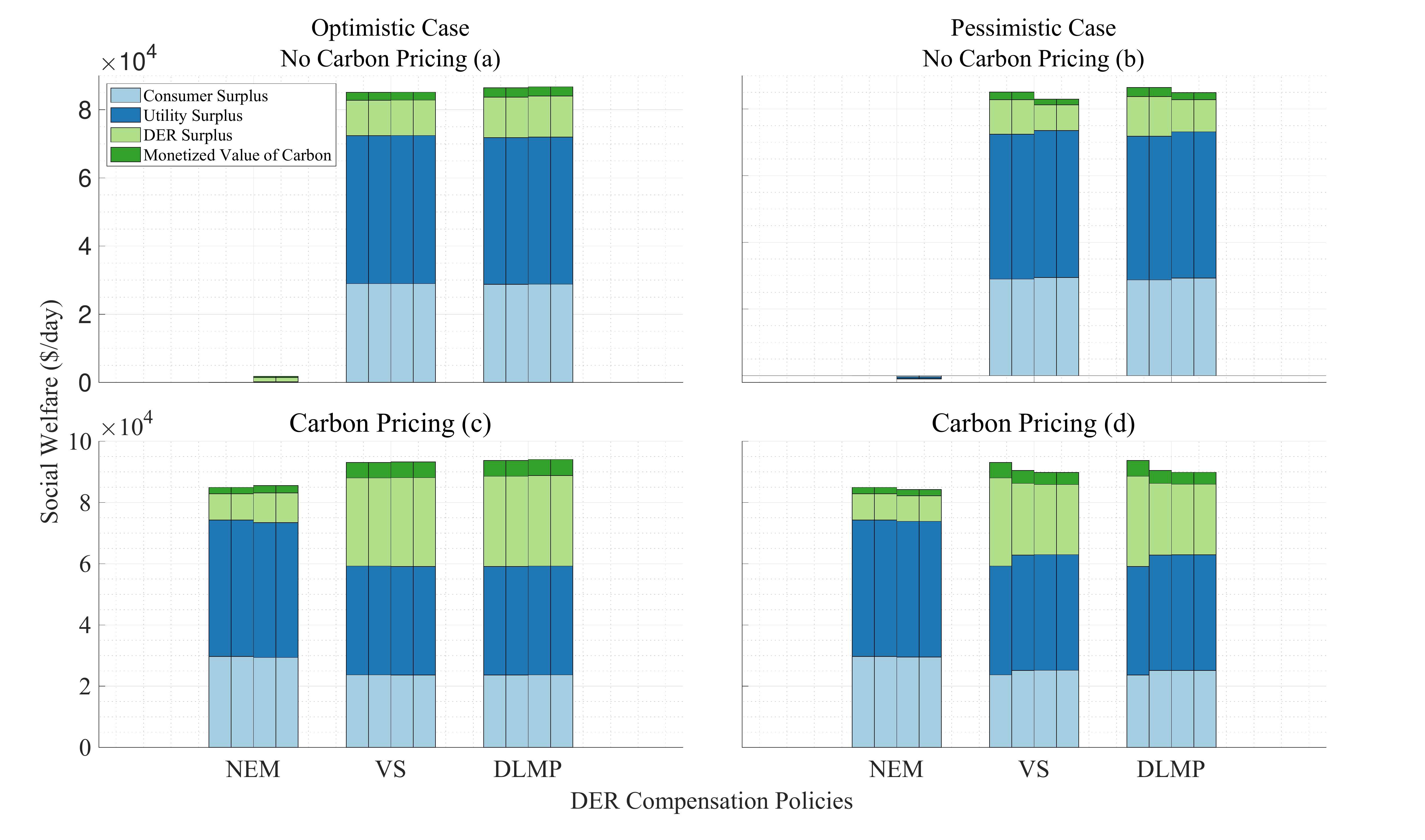}
\caption{\small{Components of system welfare incorporating market failures under different DER compensation policies. Each group of bars represents Cases 1-4 for each DER compensation policy.  (a) \& (c): Optimistic beliefs of the DER aggregator under NEM, VS, and DLMP with and without carbon pricing, (b) \& (d): Pessimistic beliefs of the DER aggregator under NEM, VS, and DLMP with and without carbon pricing.
}}
\label{obj_comp}
\end{figure}

\subsection{Effects of DER compensation policies and market failures on social welfare}

The effects of information asymmetry on social welfare under different compensation policies with and without carbon pricing are shown in Fig.~\ref{obj_comp}. We evaluate relative differences in the components of social welfare (surplus of the consumers, utility, and DER aggregator, and monetized value of carbon-dioxide emissions) for all cases, as compared to the respective components of the benchmark case. For cases with no carbon pricing, see Figs.~\ref{obj_comp}(a) and (b), NEM results in the minimum consumer surplus. Consumer surplus increases under VS and DLMP for all cases, as compared to the benchmark case, irrespective of the kind of information asymmetry or the nature of the beliefs. The increase in consumer surplus is attributed to the increase in the installed DER capacity under granular DER compensation policies, see Fig.~\ref{Cap_asymm}. Owing to higher DER capacities under VS and DLMP, the power utility procures less power from the wholesale market at LMP, reducing the costs of the power utility. The reduction in utility costs constrains the electricity tariff in the system via the utility revenue adequacy requirements, lowering the costs to the consumers, consequently increasing the consumer surplus under VS and DLMP. The trend in the differences between the monetized value of carbon-dioxide and differences in the surplus of the DER aggregator for Cases 1—4 under all the compensation policies, are identical to the trends in the installed DER capacity in Fig.~\ref{Cap_asymm}. Monetized value of carbon-dioxide and surplus of the DER aggregator increase as more DER capacity is installed, and vice versa. The surplus of the utility, being a function of the utility revenue and costs, varies with the electricity tariff, DER compensation, and LMP in the wholesale market. Under VS and DLMP, owing to higher values of DER compensation, as compared to NEM, the costs associated with DER payments increase for the power utility. However, the increase in cost is accompanied by the increase in the installed DER capacity, which allows the utility to procure less power from the wholesale market, especially during the peak load hours. This reduces the overall cost of the utility and increases the surplus of the utility under VS and DLMP as compared to NEM.

For cases with carbon pricing, see Figs.~\ref{obj_comp}(c) and (d), the trends in all the components of social welfare remain similar to the trends under no carbon pricing. However, consistent with the increase in the DER capacity in Figs.~\ref{Cap_asymm}(c) and (d), the surplus of the DER aggregator increases under all cases, see Figs.~\ref{obj_comp}(c) and (d), as compared to their counterparts under no carbon pricing, see Figs.~\ref{obj_comp}(a) and (b). The surplus of the utility under carbon pricing increases for NEM, however, decreases for VS and DLMP, as compared to respective cases with no carbon pricing. The increased surplus under NEM is due to the higher installed DER capacity which allows the utility to reduce its operational cost. However, under VS and DLMP with carbon pricing, the DER capacity increases to a point where the payments to the DER aggregators outweigh the reduction in utility cost attributed to the increase in the installed DER capacity, effectively reducing the surplus of the utility. 

Thus, information asymmetry alone minimally affect the surplus of the utility and consumers. However, the structure of DER compensation policies (flat or granular) have significant impacts on the components of the social welfare.

\section{Discussion and Conclusion}

Motivated by the most pertinent DER policy debates, this paper argues that more granular compensation policies, e.g., VS and DLMP, provide better price signals for DER deployment and improve the social welfare. We analyze the effects of information asymmetry in hosting capacity and consumer information between the power utility and the DER aggregators on the DER compensation policies, and offer the following policy- and stakeholder-relevant (e.g., public utility commissions, power utilities, and rooftop solar industry) recommendations: 


\begin{itemize}

\item Granular compensation policies provide better economic signals to DER investors for installing and operating DERs. DLMP-based compensation results in the  maximum installed DER capacity, followed by VS, and NEM, regardless of information problems.  Thus, shifting towards granular DER compensation, even in the presence of incomplete information, yields better price signals, resulting in economically efficient policies.

\item The effect of information asymmetry depends heavily on the nature of the beliefs (optimistic or pessimistic). The optimistic beliefs  either do not affect (for VS and DLMP), or benefit (for NEM) the system. 
However, pessimistic beliefs about consumer information significantly affect the market outcomes for all compensation policies. Thus, for systems where information asymmetry in consumer information might result in pessimistic beliefs, policymakers need to devise effective mitigation mechanisms for alleviating such market failures.


\item Carbon pricing  enhances incentives to install and operate DERs. When incorporated, the total installed DER capacity and social welfare increase significantly under all compensation policies, as compared to the status quo. 
We argue that carbon pricing has the potential to intrinsically offset out-of-market actions such as the currently devised incentives and subsidies to efficiently increase the penetration of DERs.

\item The overall effects of information asymmetry under any DER compensation policy are exacerbated with carbon pricing. However, the benefits achieved by incorporating carbon pricing greatly outweigh the exacerbation associated with the information problems. Therefore, policymakers should strive to introduce carbon pricing as an intrinsic market incentive, while simultaneously developing regulations (mandates for utilities to publicly share information) and incentives (e.g., performance-based regulation to condition utility profits with DER deployment targets) to facilitate access of information to all the stakeholders. Continued efforts to alleviate the information problems without carbon pricing would result in a better market performance than status quo, however, still far from an economically efficient market. 
\end{itemize}

\pagebreak

\section{Methods}

\subsection*{Notation}
\subsection*{Sets and Indices:}
\begin{tabular}{c l}
$b_0$ & Root node of the distribution network\\
$B_{m/n} (b)$ & Set of ancestor/children nodes of node $b$\\
$b^\textnormal{T}_c$&  Node in the transmission network connected to $b_0$ \\
\textbf{$b \in B$} & Set of nodes in the distribution network\\
$H_{b}^\textnormal{DER}$ & Set of hosting capacity values constituting the belief of DER aggregator at node $b$\\
$i \in I$ & Set of generators in the distribution network\\
$i \in I^{\textnormal{DER}}$ & Set of generators owned by the DER aggregator\\
$r \in R$ & Set of representative operating days\\
$T_1/T_2$ & Set of time intervals in peak/off-peak demand times \\
$T$ & Set of all time intervals ($T_1 \cup T_2$)\\
\end{tabular}

\subsection*{Parameters:}
\begin{tabular}{c l}
    
$C_{i}$ & Operational cost of generator $i\in I^{\textnormal{DER}}, I$ \\ 
$C^{\textnormal{cap}}$ & Capital cost of utility (prorated on daily basis)\\
$C_{i}^{\textnormal{inv}}$ & Prorated investment cost of generator $i \in I^{\textnormal{DER}}$   \\
$D^\textnormal{{p}}_{b,t,r}$ & Total inflexible active power demand at node $b$\\
$d^\textnormal{{q}}_{b,t,r}$ & Inflexible reactive power demand at node $b$\\
$d^\textnormal{p/q,DER}_{b,t,r}$ & Belief of DER aggregator w.r.t. active/reactive power demand at node $b$\\
$E(g_{b_o,t,r})$ & Emissions for the production of $g_{b_o,t,r}$\\
$F^{\textnormal{max}}_{(b^\textnormal{T}_c,b_o)}$ & Maximum allowable power flow in the line connecting $b^\textnormal{T}_c$ and $b_o$\\
$G^{{\textnormal{max/min}}}_{i}$ & Maximum/minimum power output of generator $i$\\
$H_{b}$ & Hosting Capacity at node ${b}$\\
$M_{b,t,r}/N$ & Utility function parameters\\
$Q^{{\textnormal{max/min}}}_{i}$ & Maximum/minimum reactive power output of generator  $i$\\
$R_{i}$ & Emission factor of generator $i$ for  carbon-dioxide emissions [ton/MWh] \\
$S_{(b,b_1)}$ & Maximum apparent power flow in distribution feeder between $b$ and $b_1 \in B$\\
$U^{{\textnormal{max/min}}}_{b}$ & Maximum/minimum square of voltage magnitude at node $b$\\
$X_{(b,b_1)}$ & Resistance of line connecting $b$ and $b_1 \in B$\\
$x_{(b,b_1)}$ & Reactance of line connecting $b$ and $b_1 \in B$\\
$\kappa_i$ & Capacity factor of generator $i \in I^\textnormal{DER}$\\
$\kappa'_{i,t,r}$ & Forecast factor of generator $i \in I^\textnormal{DER}$ at time $t$\\
$\gamma / \gamma^\textnormal{EC}$ & Penalty/ environmental cost of CO\textsubscript{2} emissions [$\$$/ton] \\
$\upsilon$ & Rate of return for power utility \\
$\phi$ & Proportionality coefficient between changes in active and reactive power demands\\

\end{tabular}

\subsection*{Variables:}
\begin{tabular} {c l}
    
$d^\textnormal{{p}}_{b,t,r}$ & Flexible active power demand at node $b$\\
$e$ & Carbon-dioxide emissions in the distribution network\\
$e_{b}$ & Carbon-dioxide emissions in the distribution network at node $b$\\
$f^{\textnormal{p/q}}_{(b,b_m),t,r}$ & Active/reactive power flow in line between $b$ and $b_m$\\
$g_{b,t,r}$ & Power output of generator connected to node $b$\\
$g_{i,t,r}$ & Power output of generator $i \in I$\\
$g_{i,t,r}^{\textnormal{T/D}}$ & Power output of generator $i \in I^{\textnormal{DER}}$ in transmission/distribution network for time interval $t$
\end{tabular}\\
\begin{tabular}{c l}
$g_{i}^{\textnormal{max}}$ & Capacity of generator $i \in I^{\textnormal{DER}}$\\
$(g/q)_{b_0,t,r}$ & Active/reactive power output of generator at $b_o$\\
$q_{b,t,r}$ & Reactive power output of generator at node $b$\\
$u_{b,t,r}$ & Square of voltage magnitude at node $b$\\
$\lambda^{\textnormal{T}}_{b^{\textnormal{T}}_c,t,r}$ & Wholesale market-clearing price at node $b^{\textnormal{T}}_c$\\
$\pi_{t,r}^\textnormal{p/op} \in \pi_{t,r}$ & Retail electricity tariff during peak/off-peak time intervals $t \in T_1$ or  $t \in T_2$\\
$\pi^{\textnormal{DER}}_{b,t}$ & DER compensation in the distribution network at node $b$\\
\end{tabular}

\subsection{Modeling Framework}
\begin{figure}[!t]
\centering
\includegraphics[scale = 0.6]{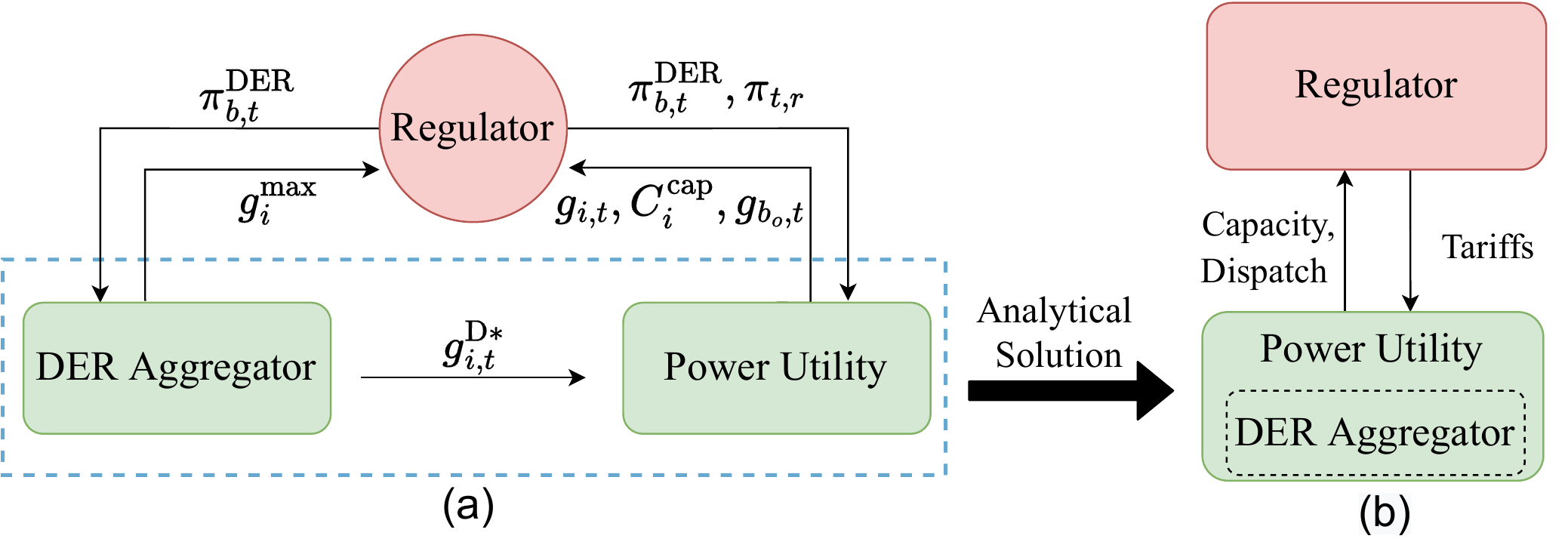}
\caption{\small{A schematic representation of the bilevel game between regulator, DER aggregator, and power utility. (a): SLMF game, (b): SLSF game}}
\label{schematic_1}
\end{figure}

Here, we use the Stackelberg game approach to model the interactions between the three key stakeholders of the DER compensation policy debate: regulator, DER aggregator, and power utility. The regulator (generally referred to as the Public Utility Commission) moves first by choosing the DER compensation and retail electricity tariff. The DER aggregator and the power utility then follow by optimizing their respective investment and operation decisions. Therefore, we model the regulator as the leader, while the DER aggregator and the power utility constitute the two followers, giving rise to a Single Leader Multi--Follower (SLMF) game, shown in Fig.~\ref{schematic_1}(a). Our model assumes that based on the upper-level (UL) decision variables (problem of the regulator), the DER aggregator maximizes its profit by optimizing its investment decisions on DER location and capacity, and the operational decisions about the share of this capacity  offered to the power utility. Simultaneously, based on the offered capacity and the UL decision variables, the power utility maximizes its regulated profit. We derive optimality conditions for the profit maximization problem of the DER aggregator in Section~\ref{DER}. These conditions are embedded into the problem of the power utility to convert the SLMF game into a Single Leader Single Follower (SLSF) game, as shown in Fig.~\ref{schematic_1}(b).

\begin{figure}[!t]
\centering
\includegraphics[scale = 0.7]{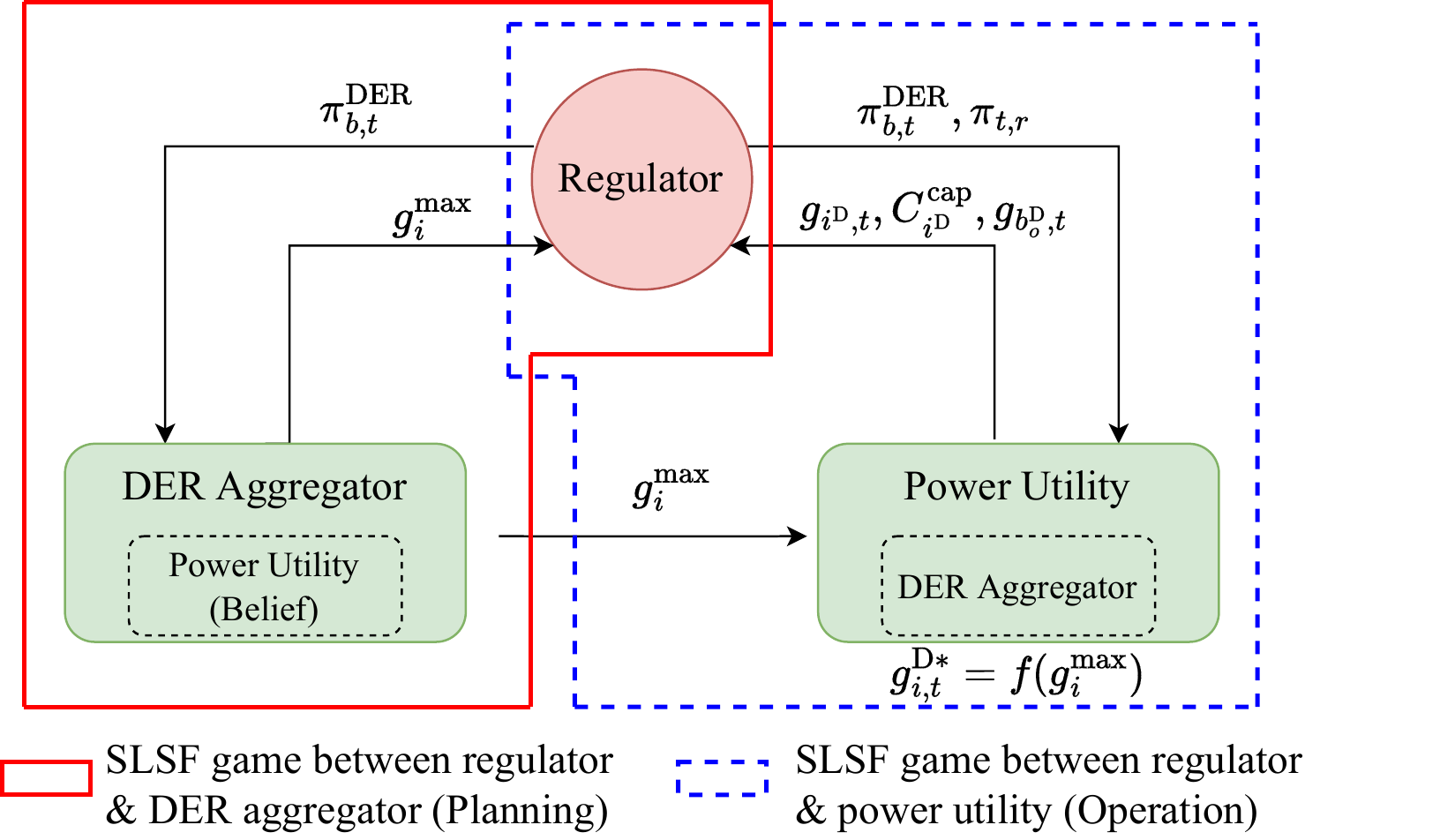}
\caption{\small{A schematic representation of the sequential SLSF game modeling information asymmetry between the DER aggregator and power utility.}}
\label{schematic}
\end{figure}

To model information asymmetry between the DER aggregator and the power utility, we re-model the SLSF game, shown in Fig.~\ref{schematic_1}(b), as two sequential SLSF games to incorporate the beliefs of the DER aggregator about the system, see Fig.~\ref{schematic}.
The regulator and the DER aggregator form the first SLSF game (shown in red in Fig.~\ref{schematic}), resulting in  optimal location and  capacity decisions for the DER aggregator. Note that these decisions are based on the belief of the DER aggregator about the distribution network, and not on the actual distribution network parameters (in case of information asymmetry between the stakeholders, the stakeholders with no or incomplete access to the system information bridge this gap by forming their own beliefs about the missing information). The optimal  decisions from the first SLSF game then parametrize the decisions of the second SLSF game, i.e., between the regulator and the power utility (shown in dotted blue in Fig.~\ref{schematic}).

The proposed problem formulation also underscores the distinct temporal nature of the two decision variables of the DER aggregator. The DER installation decisions ($g_i^{\textnormal{max}}$) form a planning problem, whereas the time--varying offer of  the DER aggregator in the distribution system ($g_{i,t,r}^{\textnormal{D}\ast}$) is a typical operational issue. Hence, the dissection of the SLMF game into two hierarchical SLSF games not only finds its motivation in terms of modeling information asymmetry, but also in the realistic timeline of decisions  involving the DER roll--out and operation. In the following sections, we describe the mathematical formulation for each player in the sequential SLSF game.

\subsubsection{Formulation of the Regulator (UL Problem)}
The regulator, modeled as the UL problem in Fig.~\ref{schematic}, aims to maximize social welfare in the system, which includes the surplus of the consumers, the surplus of the power utility, the surplus of the DER aggregators, and the environmental damage due to carbon-dioxide emissions associated with the energy production. The proposed regulator model is motivated by the real--world practice of the New York Public Service Commission \cite{NYPSC}, and is given as follows: 

\allowdisplaybreaks
\begin{subequations}
\label{reg_full}
\begin{align}
\label{reg}
\begin{split}
    &\max_{\{\pi_{t,r}, \pi_b^{\textnormal{DER}}\}}~\textnormal{O} =  \underbrace{\sum_{b \in B, t \in T, r \in R}\left[{M_{b,t,r} (d^\textnormal{p}_{b,t,r} + D^\textnormal{p}_{b,t,r}) - \frac{1}{2}N (d^{\textnormal{p}}_{b,t,r} + D^\textnormal{p}_{b,t,r})^2}\right]}_\text{Utility function of consumers} -\underbrace{\sum_{b \in B, t \in T, r \in R}(\pi_{t,r}) (d^\textnormal{p}_{b,t,r}+ D^\textnormal{p}_{b,t,r})}_\text{Cost to consumers}  \\&+ 
    \underbrace{\sum_{b\in B, t \in T, r \in R}{\pi_{t,r}(d^\textnormal{p}_{b,t,r} + D^\textnormal{p}_{b,t,r})}}_\text{Revenue of utility} - \underbrace{\sum_{i \in I,t \in T, r \in R}{C_{i}g_{i,t,r}} - \sum_{t \in T, r \in R} \lambda^{\textnormal{T}}_{b^\textnormal{T}_c,t,r} g_{b_{0},t,r}-\sum_{i \in I^{\textnormal{DER}}, t \in T, r \in R}{\pi^{\textnormal{DER}}_{b}g_{i,t,r}^{\textnormal{D}}} - C^{\textnormal{cap}}}_\text{Capital and generation/procurement cost of electricity for utility}\\&
    +\underbrace{\sum_{i \in I^\textnormal{DER},t\in T, r \in R}{\left[\lambda_{b^\textnormal{T}_c,t,r}^{\textnormal{T}} g_{i,t,r}^\textnormal{T} + \pi^{\textnormal{DER}}_{b} g_{i,t,r}^\textnormal{D}\right]}}_\text{Revenue of DER aggregator} - \underbrace{\sum_{i \in I^\textnormal{DER}}{C_i^\textnormal{inv}g_i^{\textnormal{max}}} - \sum_{i \in I^\textnormal{DER},t \in T, r \in R}{C_i( g_{i,t,r}^\textnormal{T} +  g_{i,t,r}^\textnormal{D}})}_\text{Operation \& investment cost of DER aggregator}  \\&- \underbrace{(\gamma^{\textnormal{EC}} - \gamma) \left[e + E\left(\sum_{t \in T, r \in R}{g_{b_o,t,r}}\right)\right]}_\text{Net monetized value of carbon-dioxide emissions} 
    \end{split}\\
    \label{rev_adeq}
    \begin{split}
& \sum_{b\in B, t \in T, r \in R}{\pi_{t,r}(d^\textnormal{p}_{b,t,r} + D^\textnormal{p}_{b,t,r})} = (1 + \upsilon)C^{\textnormal{cap}}  + \left[\sum_{i \in I,t \in T, r \in R}{C_{i}g_{i,t,r}} + \sum_{t \in T, r \in R} \lambda^\textnormal{T}_{b^{\textnormal{T}_c},t,r} g_{b_{0},t,r}\right]
\end{split}\\
\label{tariff_def}
&\pi_{t,r} =
\begin{cases}
\pi_{t,r}^\textnormal{p}, &\forall t \in T_1, r \in R  \\ \pi_{t,r}^{\textnormal{op}}, &\forall t \in T_2, r \in R
\end{cases}
\end{align}
\end{subequations}

The surplus of the consumers includes a quadratic utility function of consumers, explained in Section~\ref{Sec:Consumers}, along with the cost paid by the consumers. Similarly, the surplus of the power utility comprises the tariff--based revenue collected from the consumers minus the cost of power production, the cost of power procured from the wholesale market, the cost of power injected by the third-party DERs, and the prorated capital cost of the power utility. The surplus of the DER aggregator includes DER revenues, and the operational and prorated investment costs of the DER aggregator. Finally, the total system--wide carbon-dioxide emissions are weighted by their net external cost $(\gamma - \gamma^{\textnormal{EC}})$ to calculate the environmental damage of carbon-dioxide emissions. We use the net external cost instead of the complete environmental cost of carbon ($\gamma^{\textnormal{EC}}$) to avoid double-counting the environmental damage associated with carbon-dioxide emissions. The total system--wide emissions are calculated as the sum of the emissions produced by the generators in the distribution network and the emissions attributed to the interface power flow between the transmission and distribution (T\&D) networks. Eq.~\eqref{rev_adeq} models the revenue adequacy for the power utility, i.e., the revenue collected by the power utility is sufficient to recover its capital cost along with a pre-negotiated rate of return $(\upsilon)$, and break-even its operational cost.  Eq.~\eqref{tariff_def} models a relationship for the Time-of-Use (TOU) tariffs during peak and off-peak hours.

\subsection{Formulation of the DER Aggregator}
\label{DER}
The DER aggregator maximizes its profit from the participation in the wholesale market and compensation from the power utility, while minimizing the investment costs to install generators $i \in I^\textnormal{DER}$. The problem of the DER aggregator is:

\begin{subequations}
\label{DER_2}
\begin{align}
\label{1}
\begin{split}
    \max_{\Xi^{\textnormal{DER}}}&  \sum_{i \in I^\textnormal{DER}, b \in B,t \in T, r \in R}{(\lambda_{b^\textnormal{T}_c,t,r}^{\textnormal{T}} - C_{i(b)})g_{i(b^\textnormal{T}_c),t,r}^\textnormal{T}}  + \sum_{i \in I^\textnormal{DER}, b \in B,t \in T, r \in R} {(\pi^{\textnormal{DER}}_{b} - C_{i(b)})g_{i(b),t,r}^\textnormal{D}} -\sum_{i \in I^\textnormal{DER}}{C_i^\textnormal{inv}g_{i}^{\textnormal{max}}}
\end{split}\\
\text{s.t.}\hspace{10PT}
\label{3}
\begin{split}
& g_{i,t,r}^\textnormal{D}+ g_{i,t,r}^\textnormal{T} = \kappa_i \kappa'_{i,t,r} g_{i}^{\textnormal{max}}; \hspace{10pt} (\nu_i) \hspace{10pt}\forall i \in I^{\textnormal{DER}}, t \in T, r \in R
\end{split}\\
&\text{where}\hspace{5pt} \Xi^\textnormal{DER}\coloneqq \{g_{i,t,r}^\textnormal{D}, g_{i,t,r}^\textnormal{T}, g_{i}^{\textnormal{max}} \}.\nonumber
\end{align}
\end{subequations}

where the value of DER generation dispatched in the distribution network depends on the hosting capacity and the demand at the node at which the DER is connected, i.e. $g_{i,t,r}^\textnormal{D} := f(H_{b},  d^\textnormal{p}_{b,t,r})$. 
Constraint on maximum available power for each generator $i \in I^\textnormal{DER}$ is modeled in \eqref{3}.
Under first-order optimality conditions, \eqref{DER_2} is maximized if:

\begin{equation}
\label{optimality2}
g_{i(b),t,r}^{\textnormal{T}^\ast} = \frac{C_{i(b)}^{\textnormal{inv}} }{C_{i(b)}^{\textnormal{inv}} -\kappa_i \kappa'_{i,t,r}(\lambda^{\textnormal{T}}_{b^\textnormal{T}_c,t,r} - C_{i(b)})} - g_{i(b),t,r}^{\textnormal{D}^\ast}
\end{equation}

\begin{equation}
\label{optimality3}
g_{i(b),t,r}^{\textnormal{D}^\ast} = \frac{C_{i(b)}^{\textnormal{inv}} }{C_{i(b)}^{\textnormal{inv}} -\kappa_i \kappa'_{i,t,r}(\pi^{\textnormal{DER}}_{b}- C_{i(b)})} - g_{i(b),t,r}^{\textnormal{T}^\ast} : (\alpha_{i,t,r})
\end{equation}
Substituting \eqref{optimality3} in \eqref{3}, we obtain the following:
\begin{equation}
    \label{gmax_final}
    g_{i(b)}^{\textnormal{max}^\ast} \hspace{-5pt}= \hspace{-4pt} \frac{C_{i(b)}^{\textnormal{inv}}}{\kappa_i \kappa'_{i,t,r}(C_{i(b)}^{\textnormal{inv}} -\sum_{t \in T, r \in R}{\kappa_i \kappa'_{i,t,r}(\pi^{\textnormal{DER}}_{b}- C_{i(b)} ))} } : 
    (\xi_i)
\end{equation}
\subsection{Formulation of the Power Utility}
The objective function of the power utility in~\eqref{6} is to maximize its regulated profit, which is defined as the difference between the tariff-based revenue from selling electricity to consumers and the operating cost.  In turn, the operating cost includes the production cost of distribution-level generation resources,  the cost of the interface flow  between the wholesale market and the power utility, compensation to the DER aggregator, and carbon penalty. Electric power distribution is modeled using  the \textit{LinDistFlow} AC power flow approximation \cite{lindist} in (\ref{7}) -   (\ref{121}). The nodal active and reactive power balance at the root node of the distribution network are enforced in \eqref{7} and \eqref{8}, while at all other nodes in (\ref{9}) and (\ref{91}).  Similarly, capacity constraints for all generating units are modeled in (\ref{10}) and (\ref{14}). Nodal voltage and line power flow limits are enforced in (\ref{12}), \eqref{12.1},  and (\ref{13}). Constraint \eqref{12.1} models the  power exchange limit between the T\&D systems, whereas \eqref{13} constrains the apparent power  flows in the distribution lines \cite{lindist}. Note that  \eqref{13} is a conic constraint, where $K := \{x \in \mathbb{R}^3 \vert x_1^2 \geq x_2^2 +x_3^2\}$ and  $K^\ast$  denote primal and dual second-order cones \cite{boyd}.  Nodal voltage magnitudes are modeled in \eqref{121}. Eq.~\eqref{j4} calculates the carbon-dioxide emissions produced by the generators at each node of the system, whereas \eqref{em_d} converts these emissions to the system-wide carbon-dioxide emissions. Since the power utility dispatches all the power injected by the DER aggregator into the system (subject to the distribution network constraints), $g_{i,t,r}^{\textnormal{D}\ast}$ is defined as a parameter in the power utility model. The value of this parameter is calculated in the DER aggregator problem, defined in~\eqref{DER_2}.

\begin{subequations}
\label{distribution}
\begin{align}
\label{6}
\begin{split}
\max_{\Xi^\textnormal{U}} &\sum_{b\in B, t \in T, r \in R}\hspace{-5 pt}{\pi_{t,r}(d^\textnormal{p}_{b,t,r} + D^\textnormal{p}_{b,t,r}}) - \sum_{i \in I,t \in T, r \in R}\hspace{-5 pt}{C_{i}g_{i,t,r}}  -\sum_{t \in T, r \in R} \hspace{-5 pt}\lambda^{\textnormal{T}}_{b^\textnormal{T}_c,t,r} g_{b_{0},t,r} - \sum_{i \in I^{\textnormal{DER}}, t \in T, r \in R}\hspace{-5pt}{\pi^{\textnormal{DER}}_{b}g_{i,t,r}^{\textnormal{D}\ast}} - \gamma e
\end{split}\\
\text{s.t. \hspace{5pt}\{}
\label{7}
&{g_{b_{0},t,r}} = \sum_{b_n \in B_n(b_{0})}{f^p_{(b_{0},b_n),t,r}}: (\lambda_{b_0,t,r})\\
\label{8}
&{q_{b_{0},t,r}} = \sum_{b_n \in B_n(b_{0})}{f^\textnormal{q}_{(b_{0},b_n),t,r}}: (\lambda_{b_0,t,r}^\textnormal{q})\\
\label{9}
\begin{split}
&g_{b,t,r} + \sum_{i \in I^{\textnormal{DER}}}{g_{i(b),t,r}^{\textnormal{D}\ast}} + \sum_{b_m \in B_m(b)}{f^\textnormal{p}_{(b,b_m),t,r}} = d^\textnormal{p}_{b,t,r}+ D^\textnormal{p}_{b,t,r} + \sum_{b_n \in B_n(b)}{f^\textnormal{p}_{(b,b_n),t,r}} :(\lambda_{{b,t,r}}); \hspace{3mm} \forall b \in B
\end{split}\\
\label{91}
\begin{split}
&q_{b,t,r} + \sum_{b_m \in B_m(b)}{f^\textnormal{q}_{(b,b_m),t,r}} = d^\textnormal{q}_{b,t,r}+\hspace{-8pt}\sum_{b_n \in B_n(b)}{f^\textnormal{q}_{(b,b_n),t,r}}: (\lambda_{{b,t,r}}^\textnormal{q}); \hspace{5mm} \forall b \in B 
\end{split}\\
\label{10}
  &G_{i}^{{\textnormal{min}}} \leq  g_{i,t,r} \leq G_{i}^{{\textnormal{max}}}:(\underline{\delta}_{i,t,r}, \overline{\delta}_{i,t,r}); \forall i \in I\\
\label{14}
  &Q_{i}^{{\textnormal{min}}} \leq  q_{i,t,r} \leq Q_{i}^{{\textnormal{max}}}:(\underline{\theta}_{i,t,r}, \overline{\theta}_{i,t,r});\forall i \in I\\
 \label{12} 
 &U_{b}^{{\textnormal{min}}} \leq  u_{b,t,r} \leq  U_{b}^{{\textnormal{max}}}:(\underline{\mu}_{b,t,r}, \overline{\mu}_{b,t,r}); \forall b \in B \\
 \label{12.1}
 &-F_{(b^\textnormal{T}_c,b_0^\textnormal{D})}^{\textnormal{max}}\leq f_{(b^\textnormal{T}_c,b_o^D),t,r} \leq F_{(b^\textnormal{T}_c,b_0^\textnormal{D})}^{\textnormal{max}}: (\underline{\tau}_{(b^\textnormal{T}_c,b_0^\textnormal{D})}, \overline{\tau}_{(b^\textnormal{T}_c,b_0^\textnormal{D})})\\
 \label{new_capacity}
 & 0 \leq S_{(b,b_1)} \leq S_{(b,b_1)}^{\textnormal{max}}; (\underline{\tau}_{(b,b_1)}, \overline{\tau}_{(b,b_1)})\\
\label{13}
\begin{split}
&[S_{(b,b_1)}; f^\textnormal{p}_{(b,b_1),t,r}; f^\textnormal{q}_{(b,b_1),t,r}] \in K: ([\eta_{(b,b_1)}^\textnormal{s}; \eta_{(b,b_1),t,r}^\textnormal{p};\eta_{(b,b_1),t,r}^\textnormal{q}]) \in K^*; \forall b, b_1 \in B
\end{split}\\
\label{121}
\begin{split}
&\sum_{b_m \in B_m(b)}2(X_{(b,b_m)}f^\textnormal{p}_{(b,b_m),t,r} + x_{(b,b_m)}f^\textnormal{q}_{(b,b_m),t,r})+ u_{b,t,r}= \sum_{b_m \in B_m}\hspace{-5pt}u_{b_{m,t,r}}: (\beta_{(b,b_m),t,r}); \forall b \in B \} t \in T, r \in R
\end{split}\\
\label{j4}
&e_{b} =\sum_{i^{\textnormal{D}} \in I^{\textnormal{D}}(y), t \in T, r \in R}{R_{i^{\textnormal{D}}}g_{i(b),t,r}}: (\psi_{b}); \forall b \in B\\
\label{em_d}
&e = \sum_{b \in B}{e_{b}^\textnormal{D}}: (\chi)\\
&\text{where}\hspace{5pt} \Xi^\textnormal{U}\coloneqq \{d^\textnormal{p}_{b,t,r}, g_{i,t,r}, q_{i,t,r}, e^{\textnormal{D}}_{y}, g_{b_{0},t,r}, q_{b_{0},t,r},\nonumber f^\textnormal{p}_{(b,b_1),t,r}, f^\textnormal{q}_{(b,b_1),t,r}, u_{b,t,r}\}\nonumber
\end{align}
\end{subequations}

\subsection{Formulation of the Consumers}
\label{Sec:Consumers}
Consumers respond to the tariff ($\pi_{t,r}$) by adjusting their flexible demand ($d^\textnormal{p}_{b,t,r}$) to maximize their surplus. We use a quadratic utility function to model the response of the aggregated consumers to  tariff changes at each node. For the following utility function:

\begin{align}
    U(d^\textnormal{p}_{b,t,r}) =
    \begin{cases}
    M_{b,t,r} d^\textnormal{p}_{b,t,r} - \frac{1}{2}N (d^{\textnormal{p}}_{b,t,r})^2 & \hspace{-5pt}; \text{if}~ 0 \leq d^\textnormal{p}_{b,t,r} \leq \frac{M_{b,t,r}}{N}\\
    \frac{M_{b,t,r}^2}{2N} & \hspace{-5pt}; \text{if}~ d^\textnormal{p}_{b,t,r} \geq \frac{M_{b,t,r}}{N}
    \end{cases}
\end{align}
we define the consumer surplus as:

\begin{align}
\label{CS}
   \text{CS}:=  \hspace{-10pt}&\sum_{b \in B, t \in T, r \in R}{ \underbrace{M_{b,t,r} d^\textnormal{p}_{b,t,r} - \frac{1}{2}N (d^{\textnormal{p}}_{b,t,r})^2}_\text{Utility function of Consumers} - \underbrace{\pi_{t,r} d^\textnormal{p}_{b,t,r}}_\text{Cost to consumers}}
\end{align}

Now, for a non-zero marginal utility of electricity, we derive an analytical solution for demand flexibility as a function of the electricity tariff. Using the first-order optimality conditions of \eqref{CS}, we obtain:

\begin{align}
\label{demand_flex}
    d^\textnormal{p}_{b,t,r}(\pi_{t,r}) = \frac{M_{b,t,r}-\pi_{t,r}}{N}
\end{align}
Hence, given a particular electricity tariff set by the regulator, the consumers would adjust their demand as in \eqref{demand_flex}.

\subsection{Lower-Level (LL) Problem}
To obtain the LL problem, we embed the DER aggregator problem  in~\eqref{DER_2} into the power utility formulation in~\eqref{distribution}, using the first-order optimality results in~\eqref{optimality2}-\eqref{gmax_final}. The resulting problem is as follows:
\begin{subequations}
\label{LL Problem}
\begin{align}
\label{LL_Problem_1}
    \text{max}: \hspace{10pt}& \text{Eq.}~\eqref{6}\\
    \label{LL_Problem_2}
    \text{s.t.} \hspace{10pt}& \text{Eqs.}~\eqref{3}, \eqref{optimality2},\eqref{gmax_final}, \eqref{7} - \eqref{em_d}\\
    \label{LL_3}
    & g_{i(b)}^{\textnormal{max}} \leq H_{b}; \forall b \in B : (\epsilon_i)
\end{align}
\end{subequations}
where~\eqref{LL_3} constrains the maximum DER capacity of $i \in I^\textnormal{DER}$ that can be installed at node $b$ to its respective hosting capacity. Hence, using the analytical solution of the problem of the DER aggregator, we convert the LL problem from two players to a single player, as shown in Fig.~\ref{schematic_1}.

\subsection{Modeling DER Compensation Policies} \label{Sec:Rem_Pol}
This section describes and formulates the three DER compensation policies (Net Energy Metering (NEM), Value Stack (VS), and Distributional Locational Marginal Price (DLMP)) considered in this paper. The resulting equations define $\pi_b^\textnormal{DER}$ in the UL problem (eq.~\eqref{reg_full}) and the LL problem (eq.\eqref{LL Problem}), to analyze the efficiency impacts of the aforementioned DER compensation policies.

\subsubsection{NEM-Based Compensation}
NEM is the most common compensation mechanism used by the utilities to compensate DERs for behind-the-meter generation. The traditional NEM policy is functionally equivalent to a single electricity meter that can either run in the forward or the backward direction. When the energy production by the customer is less than their demand, the meter runs in the forward direction, and in the backward direction, otherwise. Customers are billed for the net energy consumption at the end of the billing cycle, using the retail electricity tariff \cite{Burcin_NEM}. Thus, we define:

\begin{align} \label{NEM}
    \pi^{\textnormal{NEM}} \coloneqq \pi^{\textnormal{DER}}_b = \pi_{t,r}
 \end{align}
 and integrate it with the UL problem in~\eqref{reg_full} to formulate the NEM policy for DER compensation.
 
 \subsubsection{VS-Based Compensation}
VS compensation depends on the quantification of benefits derived from individual attributes of a commodity. In case of DERs, the value of the commodity (injected power into the gird) is derived from its temporal and locational attributes, i.e., the time and location of the DER power injection in the grid. The determination of these attributes varies across jurisdictions, however, in New York the VS includes  energy, capacity, environmental, demand reduction, and locational system relief values of DER \cite{VDER_1}. Thus, DERs receive compensation for the energy and capacity purchase requirements of the utility offset by the DER injection, and average distribution network costs avoided by the DER injections. Moreover, DER injections in specific areas (predetermined by the utility) are additionally compensated for easing network constraints, e.g., congestion. Accordingly, we define the VS-based compensation as:

\begin{align} \label{ValueStack}
    \pi^{\textnormal{VS}} \coloneqq \pi^{\textnormal{DER}}_b = \lambda^{\textnormal{T}}_{b^\textnormal{T}_c,t,r} + \gamma^{\textnormal{EC}}R_{i} + \overline{\tau}_{(b,b_1)}
\end{align}
where the LMP ($\lambda^{\textnormal{T}}_{b^\textnormal{T}_c,t,r}$) at the interconnecting node between the wholesale market and the distribution network is used to model the avoided energy costs, external costs of CO\textsubscript{2} emissions from the marginal generator at time $t$ for day $r$ are used to model the environmental value, and avoided capacity and network costs are represented by $\overline{\tau}_{(b,b_1)}$, which is the dual variable of the capacity constraint \eqref{new_capacity}. 

\subsubsection{DLMP-Based Compensation} \label{DLMP}
In case of a deregulated distribution system, a market operator solves the optimal power flow to minimize costs at the distribution level. Similar to the existing wholesale market, the market clearing price is then used to compensate DERs. The market clearing price is DLMP, and is calculated using the dual variable of the active power balance constraint in \eqref{9}. Thus, the DLMP-based DER compensation is:

\begin{align} \label{DLMP}
 \pi^{\textnormal{DLMP}} \coloneqq \pi^{\textnormal{DER}}_b = \lambda_{b,t,r}
\end{align}

\subsection{Modeling Information Asymmetry}
DER-specific information problems can originate either from the stochastic variables in electricity markets (solar and wind generation on a particular day) or from the propriety information of different stakeholders \cite{Burcin_IAEE}. While the former type of information problems can never be completely resolved and leads to limited competitive advantage \cite{Robert_pricing}, the latter type not only creates  competitive advantages in electric power distribution, but also generates incentives for the stakeholders to extract information rent based on this advantage \cite{Burcin_IAEE}. Hence, in this paper, we focus on the information asymmetry between the power utilities and the DER aggregators due to the private information of stakeholders, such as distribution network information and consumer data for utilities, and DER characteristics for DER developers/aggregators.

\subsubsection{Modeling Information Asymmetry in the LL Problem:}
\subsubsection*{Information Asymmetry in the Hosting Capacity:}
\label{Sec:Hosting_Cap}
We model information asymmetry in hosting capacity between the DER aggregator and the power utility using the actual hosting capacity of the node where  DERs have to be connected (known to the power utility), and a set of values that constitute the belief of the DER aggregator about the hosting capacity. The difference between the actual value of hosting capacity and the belief of DER aggregator about this value constitutes information asymmetry. To incorporate information asymmetry in the LL problem, defined in~\eqref{LL Problem}, we re-define~\eqref{LL_3} as:

\begin{equation}
\label{hosting_info}
    g_{i(b)}^{\textnormal{max}} \leq h_{b}^\textnormal{DER}; \forall b \in B, h_{b}^\textnormal{DER} \in H_{b}^\textnormal{DER}
\end{equation}
where $H_{b}^\textnormal{DER}$ is the set of values constituting the belief of the DER aggregator. In case $H_{b} = h_{b}^\textnormal{DER} \in H_{b}^\textnormal{DER}$, there is no information asymmetry in the problem, however, as  $h_{b}^\textnormal{DER}$ deviates from $H_{b}$, we expect to see the effects of information asymmetry  on the DER aggregator's investment decisions, optimal value of the DER compensation policy, and system welfare. Now, the LL problem with information asymmetry in the hosting capacity is as follows:

\begin{subequations}
\label{LL Problem_HC}
\begin{align}
    \text{max}: \hspace{10pt}& \text{Eq.}~\eqref{6}\\
    \text{s.t.} \hspace{10pt}& \text{Eqs.}~\eqref{3}, \eqref{optimality2},\eqref{gmax_final}, \eqref{7} - \eqref{em_d}, \eqref{hosting_info}
\end{align}
\end{subequations}

\subsubsection*{Information Asymmetry in the Consumer Information:}

We model information asymmetry in the granular power consumption data between the power utility and the DER aggregator such that the belief of the  DER aggregator about the nodal demand varies by a factor ($\alpha$) relative to the actual  demand. If $\alpha = 1$, the DER aggregator has complete information about the demand. However, if $\alpha \neq 1$, there is information asymmetry between the power utility and the DER aggregator. To incorporate this information asymmetry in the LL problem in~\eqref{LL Problem}, we re-define~\eqref{9} and \eqref{91} as:

\begin{align}
    \label{DER_P_1}
\begin{split}
&g_{b,t,r} + \sum_{i \in I^{\textnormal{DER}}}{g_{i(b),t,r}^{\textnormal{D}\ast}} + \sum_{b_m \in B_m(b)}{f^\textnormal{p}_{(b,b_m),t,r}} = d^\textnormal{p,DER}_{b,t,r}+ \sum_{b_n \in B_n(b)}{f^\textnormal{p}_{(b,b_n),t,r}} :(\lambda_{{b,t,r}}); \hspace{3mm} \forall b \in B, t \in T, r \in R
\end{split}\\
\label{DER_q_1}
\begin{split}
&q_{b,t,r} + \sum_{b_m \in B_m(b)}{f^\textnormal{q}_{(b,b_m),t,r}} = d^\textnormal{q,DER}_{b,t,r}+ \sum_{b_n \in B_n(b)}{f^\textnormal{q}_{(b,b_n),t,r}}: (\lambda_{{b,t,r}}^\textnormal{q}); \hspace{5mm} \forall b \in B, t \in T, r \in R 
\end{split}\\
\label{def_1}
&\hspace{50pt}d^\textnormal{p,DER}_{b,t,r} = \alpha d^\textnormal{p}_{b,t,r}\\
\label{def_2}
&\hspace{50pt}d^\textnormal{q,DER}_{b,t,r} = \phi_{(b)} d^\textnormal{p,DER}_{b,t,r}
\end{align}
Here, we model both the information asymmetry in active power and reactive power demand, but assume that the latter varies proportionally to the former, with a constant parameter $\phi$. This assumption is motivated by the power factor relating the active and reactive power demand at each node.

Hence, the LL problem with information asymmetry in consumer information can be written as follows:

\begin{subequations}
\label{LL Problem_CD}
\begin{align}
    \text{max}: \hspace{10pt}& \text{Eq.}~\eqref{6}\\
    \text{s.t.} \hspace{10pt}& \text{Eqs.}~\eqref{3}, \eqref{optimality2},\eqref{gmax_final}, \eqref{7}, \eqref{8}, \eqref{10} - \eqref{em_d}, \eqref{DER_P_1} - \eqref{def_2}
\end{align}
\end{subequations}

\subsubsection*{Information Asymmetry in Hosting Capacity and Consumer Information:}
To formulate a comprehensive model that simultaneously accounts for both the information asymmetry in hosting capacity and consumer information, we re-define the LL problem in \eqref{LL Problem} as follows:

\begin{subequations}
\label{LL Problem_Asym_2}
\begin{align}
    \text{max}: \hspace{10pt}& \text{Eq.}~\eqref{6}\\
    \begin{split}
    \text{s.t.} \hspace{10pt}& \text{Eqs.}~\eqref{3}, \eqref{optimality2},\eqref{gmax_final}, \eqref{7}, \eqref{8}, \eqref{10} - \eqref{em_d}, \eqref{hosting_info}, \\&\hspace{23pt} \eqref{DER_P_1} - \eqref{def_2}
    \end{split}
\end{align}
\end{subequations}

\subsection{SLSF Game Between the Regulator and DER Aggregator}
Given \eqref{LL Problem_Asym_2}, the first SLSF game between the regulator and the DER aggregator, shown in Fig.~\ref{schematic}, is formulated as follows:
\begin{subequations}
\label{SLSF_1}
\begin{align}
    \max_{\Xi^\textnormal{O}, \Xi^\textnormal{U}, \Xi^\textnormal{DER}} \hspace{10pt} & \text{Eq.}~\eqref{reg}\\
    \text{s.t.} \hspace{10pt} & \text{Eqs.}~\eqref{rev_adeq}, \eqref{tariff_def}\\
    \label{cases_eq}
    & g_i^{\textnormal{max}} \in \textnormal{arg}\{\text{Eq.}~\eqref{LL Problem_Asym_2}\}
\end{align}
\end{subequations}
\textbf{Remark 1:} For the case of complete information, \eqref{cases_eq} will be modified as $g_i^{\textnormal{max}} \in \textnormal{arg}\{\text{Eq.}~\eqref{LL Problem}\}$. Similarly, for the cases modeling information asymmetry individually in hosting capacity and consumer information, \eqref{cases_eq} will be replaced by $g_i^{\textnormal{max}} \in \textnormal{arg}\{\text{Eq.}~\eqref{LL Problem_HC}\}$ and $g_i^{\textnormal{max}} \in \textnormal{arg}\{\text{Eq.}~\eqref{LL Problem_CD}\}$, respectively.

\subsection{SLSF Game Between the Regulator and Power Utility}
The second SLSF game between the regulator and the power utility, shown in Fig.~\ref{schematic}, is formulated as follows:
\begin{subequations}
\label{SLSF_2}
\begin{align}
    \max_{\Xi^\textnormal{O}, \Xi^\textnormal{U}, \Xi^\textnormal{DER}/g_i^{\textnormal{max}}}: \hspace{10pt} & \text{Eq.}~\eqref{reg}\\
    \text{s.t.} \hspace{10pt} & \text{Eqs.}~\eqref{rev_adeq}, \eqref{tariff_def}\\
    & g_{i,t,r}^{\textnormal{D}\ast} \in \textnormal{arg}\{\text{Eqs.}~\eqref{LL_Problem_1} - \eqref{LL_Problem_2}\\
    \label{SLSF_2_INEQ}
    &\hspace{20pt}g_{i,t,r}^{\textnormal{D}\ast} \leq g_i^{\textnormal{max}\ast} : (\varphi_{i,t,r})\}
\end{align}
\end{subequations}
where $g_i^{\textnormal{max}\ast}$ is the optimal DER capacity from the first SLSF game. 

\textbf{Remark 2}: To incorporate the effect of DER compensation policies on the outcomes of information asymmetry between the DER aggregator and the power utility, we analyze each of the four cases defined in Remark 1 for the three DER compensation policies detailed in Sec.~\ref{Sec:Rem_Pol}. Thus, along with the modifications in Remark 1, eqs.~\eqref{NEM} - \eqref{DLMP} will be individually incorporated in eq.~\eqref{SLSF_1} to formulate the twelve cases of the effect of DER compensation policies on information asymmetry. Furthermore, eqs.~\eqref{NEM} - \eqref{DLMP} will also be be included individually in eq.~\eqref{SLSF_2}, depending on the compensation policy fixed in one of the cases in Remark 1. 

\textbf{Remark 3}: To analyze the effects of carbon pricing in the wholesale market on the market outcomes (values of DER compensation, installed DER capacity, system welfare), all the cases defined in Remark 2 are simulated with and without carbon pricing in the LMP ($\lambda^\textnormal{T}_{b^\textnormal{T}_c,t,r}$) at the interconnecting node between the wholesale market and the distribution network.

\section{Solution Methodology}
\label{Sec:Sol_Meth}
Since the proposed sequential SLSF game in Fig.~\ref{schematic} is a bilevel optimization problem, it cannot be solved using off-the-shelf solvers, and require reformulation to a single-level equivalent. Duality theory and KKT conditions are the two primary approaches to convert multi-level optimization problems into single-level equivalents \cite{conejo}, however, the former approach introduces extensive nonlinearities in the reformulation \cite{MILP}. To avoid recasting the highly nonlinear problem 
(NLP) into a mixed-integer NLP, we use KKT conditions in this paper to reformulate the SLSF games into their single-level equivalents. The KKT conditions introduce complementarity conditions in the single-level problem, yielding a Mathematical Problem with Equilibrium Constraints (MPEC) \cite{mpec_book}. MPECs are NLPs that do not satisfy the standard Mangasarian-Fromovitz constraint qualification at any feasible point, thus invalidating the standard convergence assumptions of NLPs \cite{2_kanzow}. Therefore, multiple weaker stationarity concepts are defined for MPECs, such as C, M, B, and strong stationary points \cite{relaxation_comp}. The convergence of MPECs to one of these points defines the proximity of the solution to the actual optimizer \cite{inexact}. 

Multiple solution approaches to solve MPECs are discussed in the current literature \cite{SLMF}. However, to exploit the ease of solving MPECs using off-the-shelf NLP solvers, we use the relaxation-based NLP reformulation. Although many relaxation approaches for MPECs are proposed \cite{Scholtes, lin, kadrani,ulbrich,jean}, these approaches numerically converge to an inexact and weaker stationarity result \cite{inexact}. However, the relaxation method by Scholtes \cite{Scholtes} is guaranteed to converge to a C-stationary point even if the numerical convergence is obtained to an inexact point \cite{inexact}, rendering this technique superior to its counterparts. Therefore, we use Scholtes's relaxation scheme to solve MPECs in this paper. 

\subsection{Solution Technique}
This section presents a reformulation of the bilevel problems  in \eqref{SLSF_1} and \eqref{SLSF_2} into their single--level equivalents to make them suitable for using off--the--shelf NLP solvers. We underscore that the single--level equivalents constitute MPECs, and provide Scholtes's relaxation technique for solving MPECs.
\subsubsection{KKT conditions of the LL problem:}
We reformulate the bilevel problem in \eqref{SLSF_1} into a single-level equivalent using the KKT conditions of the LL problem in \eqref{cases_eq}. The KKT conditions are given as follows: 

\allowdisplaybreaks
\begin{subequations}
\label{KKT_dist}
\begin{align} 
\label{d1}
\begin{split}
&\{(g_{b_{0},t,r}):-\lambda^\textnormal{T}_{t,r}+\lambda_{b_0,t,r} -\underline{\tau}_{(b^\textnormal{T}_c,b_0^\textnormal{D}),t,r}+\overline{\tau}_{(b^\textnormal{T}_c,b_0^\textnormal{D}),t,r} - \zeta_{b_0,t,r}=0; 
\end{split}\\
\label{d2}
&(q_{b_{0},t,r}):\lambda_{b_0,t,r}^\textnormal{q} =0;\\
\label{d5}
&(q_{i,t,r}):-\lambda_{{b(i),t,r}}^{\textnormal{q}}-\underline{\theta}_{i,t,r}+\overline{\theta}_{i,t,r} = 0;\forall i \in I\\
\label{d3}
\begin{split}
&(g_{i,t,r}):-C_{i}+\lambda_{{b(i),t,r}}-\underline{\delta}_{i,t,r}+\overline{\delta}_{i,t,r} - \sum_{y \in \theta}{R_{i}\psi_{b(i)}} = 0;\forall i \in I
\end{split}\\
\label{d6}
\begin{split}
&(f^\textnormal{p}_{(b,b_m),t,r}):-\sum_{b_n \in B_n(b)}{\lambda_{{b_n,t,r}}}+\lambda_{{b,t,r}}- 2\beta_{(b,b_m),t,r}X_{(b,b_m)}-\eta_{(b,b_m),t,r}^\textnormal{p} = 0;\forall b \in B\\
\end{split}\\
\label{d7}
\begin{split}
&(f^\textnormal{q}_{(b,b_m),t,r}):-\sum_{b_n \in B_n(b)}\lambda_{{b_n,t,r}}^{\textnormal{q}}+\lambda_{{b,t,r}}^{\textnormal{q}}-2\beta_{(b,b_m),t,r}x_{(b,b_m)}-\eta_{(b,b_m),t,r}^\textnormal{q} = 0;\forall b \in B\\
\end{split}\\
\label{d8}
\begin{split}
&(u_{b,t,r}):-\underline{\mu}_{b,t,r}+\overline{\mu}_{b,t,r}+ \sum_{b_m \in B_m(b)}\beta_{(b,b_m),t,r}-\beta_{(b,b_n),t,r}=0;\forall b \in B \};\hspace{5PT}t \in T, r \in R
\end{split}\\
\begin{split}
&(g_{i,t,r}^\textnormal{D}): -\pi_b^\textnormal{DER} + {\lambda_{b,t,r}} + \nu_i  + {\alpha_{i,t,r}} = 0; \forall i \in I^\textnormal{DER}, t \in T, r \in R 
\end{split}\\
&(e_{b}):-\psi_{b}+\chi =0; \forall b \in B\\
&(e):-\gamma- \chi =0; \\
\label{flex_dem}
&(d_{b,t,r}^\textnormal{p}): \pi_{t,r} - \alpha\lambda_{b,t,r} = 0; \hspace{10pt} \forall b \in B, t \in T, r \in R\\
\label{flex_dem_q}
&(d_{b,t,r}^\textnormal{q}):  \alpha\phi_i\lambda_{b,t,r}^{\textnormal{q}} = 0; \hspace{10pt}\forall b \in B, t \in T, r \in R\\
\label{g_i_max}
&(g_i^{\textnormal{max}}): -\sum_{t \in T, r \in R}{\kappa_i \kappa'_{i,t,r}\nu_{i}} + \xi_i - \epsilon_i; \hspace{10pt}\forall i \in I^\textnormal{DER}\\
\label{d9}
&0 \leq g_{i,t,r}-G_{{i}}^\textnormal{{min}} \;\bot\; \underline{\delta}_{i,t,r} \geq 0;\forall i \in I, t \in T, r \in R\\
\label{d10}
&0 \leq G_{{i}}^\textnormal{{max}}-g_{i,t,r} \;\bot\; \overline{\delta}_{i,t,r} \geq 0;\forall i \in I, t \in T, r \in R\\
\label{d11}
&0 \leq q_{i,t,r}-Q_{{i}}^\textnormal{{min}} \;\bot\; \underline{\theta}_{i,t,r} \geq 0;\forall i \in I, t \in T, r \in R\\
\label{d12}
&0 \leq Q_{{i}}^\textnormal{{max}}-q_{i,t,r} \;\bot\; \overline{\theta}_{i,t,r} \geq 0;\forall i \in I, t \in T, r \in R\\
\label{d13}
&0 \leq u_{b,t,r}-U_{{b}}^\textnormal{{min}} \;\bot\; \underline{\mu}_{b,t,r} \geq 0;\forall b \in B, t \in T, r \in R\\
\label{d14}
&0 \leq U_{{b}}^\textnormal{{max}}-u_{b,t,r} \;\bot\; \overline{\mu}_{b,t,r} \geq 0;\forall b \in B, t \in T, r \in R\\
\label{d14.1}
&0 \leq g_{b_{0},t,r} \;\bot\;\underline{\tau}_{(b^\textnormal{T}_c,b_0^\textnormal{D}),t,r} \geq 0; \forall t \in T, r \in R\\ 
\label{d14.2}
&0 \leq F_{(b^\textnormal{T}_c,b_0^\textnormal{D})}^{\textnormal{max}} -g_{b_{0},t,r} \;\bot\; \overline{\tau}_{(b^\textnormal{T}_c,b_0^\textnormal{D}),t,r} \geq 0; \forall t \in T, r \in R\\
\label{asymm_ineq}
&0 \leq h_b^{\textnormal{DER}} - g_{i(b)}^{\textnormal{max}}\;\bot\; \epsilon_i \geq 0; \forall b \in B, h_b^{\textnormal{DER}} \in H_b^{\textnormal{DER}}\\
\label{cone}
\begin{split}
&[S_{(b,b_1)}; f^\textnormal{p}_{(b,b_1),t,r}; f^\textnormal{q}_{(b,b_1),t,r}] \;\bot\;[\eta_{(b,b_1)}^\textnormal{s}; \eta_{(b,b_1),t,r}^\textnormal{p};\eta_{(b,b_1),t,r}^\textnormal{q}]; \forall b, b_1 \in B, t \in T, r \in R
\end{split}\\
\label{d15}
\begin{split}
&\text{Equality Constraints: Eqs.}~ \eqref{3}, \eqref{optimality2}, \eqref{gmax_final},  \eqref{7}, \eqref{8}, \eqref{121}-\eqref{em_d}, \eqref{hosting_info}, \eqref{DER_P_1}-\eqref{def_2}.
\end{split}
\end{align}
\end{subequations}
\textbf{Remark 4:} For the case of complete information, \eqref{flex_dem} will be replaced by 
\begin{equation}
\label{KKT_perf_inf}
    \pi_{t,r} - \lambda_{b,t,r} = 0; \hspace{10pt}\forall b \in B, t \in T, r \in R
\end{equation} 
whereas \eqref{flex_dem_q} will be replaced by 
\begin{equation}
\label{KKT_perf_inf_q}
    \lambda_{b,t,r}^{\textnormal{q}} = 0; \hspace{20pt}\forall b \in B, t \in T, r \in R.
\end{equation}
Moreover, \eqref{asymm_ineq} will be reformulated as 
\begin{equation}
\label{ineq_perf}
    0 \leq H_b - g_{i(b)}^{\textnormal{max}}\;\bot\; \epsilon_i \geq 0; \hspace{10pt}\forall b \in B\\. 
\end{equation}
Similarly, for the case modeling information asymmetry in the  hosting capacity, \eqref{KKT_perf_inf} and \eqref{KKT_perf_inf_q} will be used instead of \eqref{flex_dem} and \eqref{flex_dem_q}. Moreover, the equality constraints will include \eqref{9} and \eqref{91} instead of \eqref{DER_P_1} -- \eqref{def_2}. For the case modeling information asymmetry in consumer information, \eqref{ineq_perf} will be used instead of \eqref{asymm_ineq}. \\
\textbf{Remark 5:} For the parametrized SLSF game in \eqref{SLSF_2}, the KKT conditions will be the same as that of the complete information case in Remark 4, however, $0\leq g_i^{\textnormal{max}\ast} - g_{i,t,r}^{\textnormal{D}\ast}\;\bot\;\varphi_{i,t,r}\geq 0~ \forall i \in I^\textnormal{DER}, t \in T, r \in R,$ and \eqref{LL_3} will be used instead of \eqref{ineq_perf} and \eqref{g_i_max}.

\subsubsection{MPEC Reformulation:}
To obtain a single-level MPEC of the bilevel game in~\eqref{SLSF_1}, we incorporate the UL problem in \eqref{reg_full} and the KKT conditions of the LL problem in \eqref{KKT_dist} as follows: 

\subsubsection*{MPEC reformulation for the first SLSF game:}
\label{MPEC_1}
The MPEC reformulation for the SLSF game between the regulator and the DER aggregator in \eqref{SLSF_1} is given below:

\begin{subequations}
\label{MPEC_1}
\begin{align}
\label{jm}
\max_{\Xi^\textnormal{O}, \Xi^\textnormal{U}, \Xi^\textnormal{DER}}: \; \hspace{20pt}&\textnormal{Eq.}~\eqref{reg}\\
\textnormal{s.t.}\hspace{20pt}
&\text{UL Constraint:} \hspace{5pt} \textnormal{Eq.}~\eqref{rev_adeq}, \eqref{tariff_def}\\  
&\text{LL KKT Conditions:} \hspace{5pt} \textnormal{Eqs.}~\eqref{d1}-\eqref{d15} 
\end{align}
\end{subequations}

\subsubsection*{MPEC reformulation for the parametrized SLSF Game:}
\label{MPEC_2}
The MPEC reformulation for the parametrized SLSF game between the regulator and the power utility in \eqref{SLSF_2} is given as:

\begin{subequations}
\label{MPEC_2}
\begin{align}
\max_{\Xi^\textnormal{O}, \Xi^\textnormal{U}, \Xi^\textnormal{DER}/g_i^{\textnormal{max}}}: \; \hspace{20pt}&\textnormal{Eq.}~\eqref{reg}\\
\textnormal{s.t.}\hspace{20pt}
&\text{UL Constraint:} \hspace{5pt} \textnormal{Eq.}~\eqref{rev_adeq} - \eqref{tariff_def}\\  
&\text{LL KKT Conditions:} \hspace{5pt} \textnormal{Remark 4}
\end{align}
\end{subequations}

\subsection{Scholtes's Relaxation Technique}

The MPECs defined in eqs.~\eqref{MPEC_1} and \eqref{MPEC_2} do not satisfy  the Linear Independence and Mangasarian--Fromovitz constraint qualifications due to the presence of complementarity constraints. Therefore, dedicated solution techniques are required so that the stationarity concepts of MPECs, explained in Sec.~\ref{Sec:Sol_Meth}, can be applied using off-the-shelf NLP solvers. In this paper, we use the global relaxation scheme by Scholtes, which is elaborated by defining a generic MPEC of the form:

\begin{subequations} \label{eq:gen_mpec}
\begin{align}
\max_{x,y,k} \hspace{15pt}&f(x,y)\\
\textnormal{s.t.}\hspace{15pt}
&g(x,y) \geq 0\\
&h(x,y) \geq 0 \; \text{and} \;k \geq 0\\
&k^\textnormal{T} h(x,y) = 0 \label{comp}
\end{align}
\end{subequations}

\begin{figure}[!t]
\centering
\includegraphics[scale = 0.85]{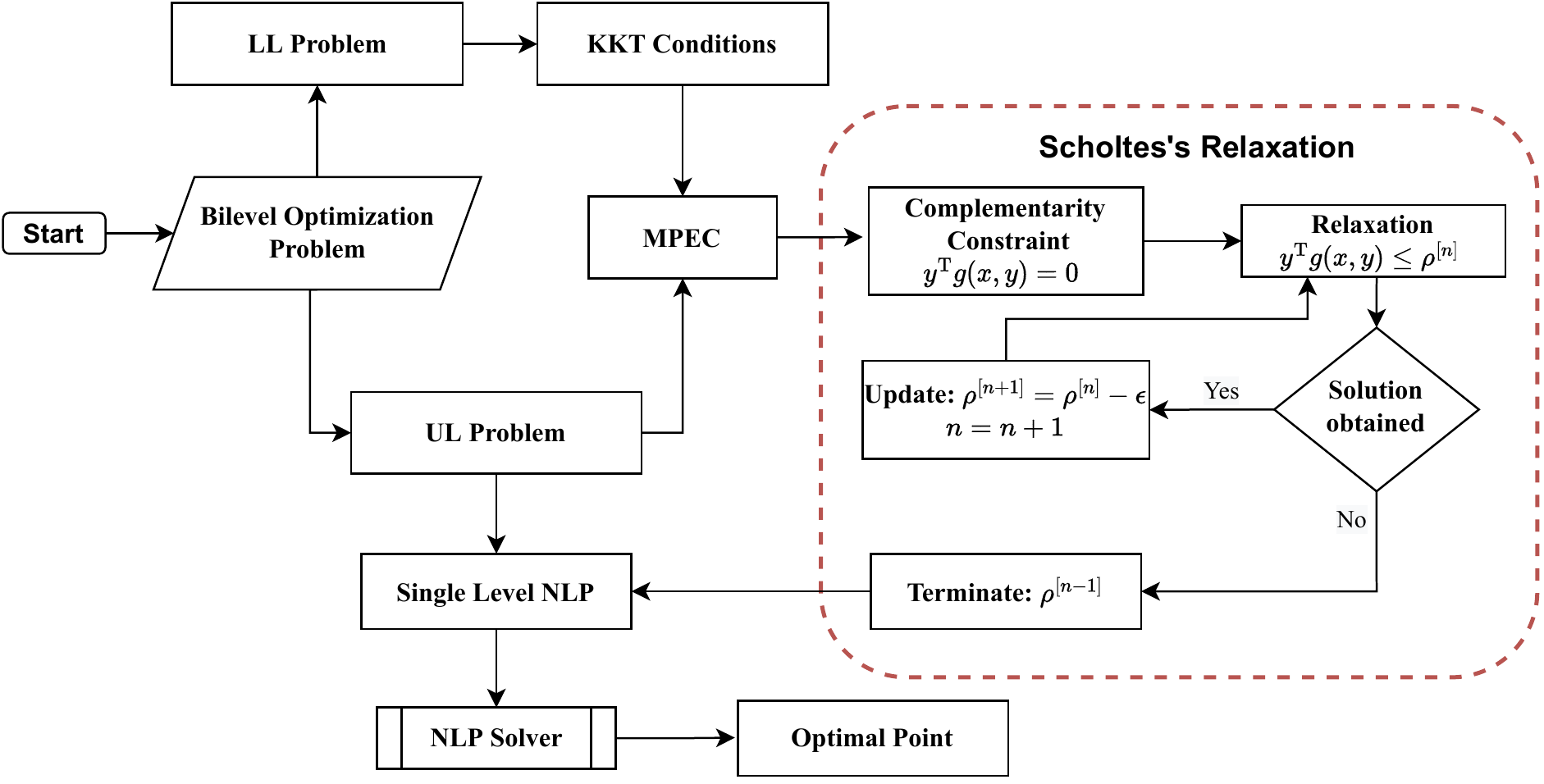}
\caption{An illustrative diagram showing the solution methodology for solving the bilevel optimization problem. It includes the reformulation of bilevel optimization into a single-level equivalent, and the use of Scholtes's relaxation technique.}
\label{scholtes}
\end{figure}

where \eqref{eq:gen_mpec}  accommodates the MPECs in eqs.~\eqref{MPEC_1} and \eqref{MPEC_2}. Scholtes' relaxation methodology relaxes the complementarity constraint, eq. (\ref{comp}), using a small user-defined relaxation value $\rho$, such that $k^\textnormal{T} h(x,y) \leq \rho$. The resulting NLP formulation is iteratively solved such that in each iteration, the value of $\rho$ is decreased from its previous value till $\rho \rightarrow 0$. The associated solution of the relaxed NLP where $\rho \approx 0$ is the solution of the MPEC. An illustration for the proposed solution methodology is shown in Fig.~\ref{scholtes}. We note that the obtained solution is the strongest possible stationarity result (C-stationary point) for MPECs \cite{inexact}.

\bibliographystyle{naturemag}
\bibliography{references}{}

\begin{thebibliography}{10}
\expandafter\ifx\csname url\endcsname\relax
  \def\url#1{\texttt{#1}}\fi
\expandafter\ifx\csname urlprefix\endcsname\relax\def\urlprefix{URL }\fi
\providecommand{\bibinfo}[2]{#2}
\providecommand{\eprint}[2][]{\url{#2}}

\bibitem{DSIRE}
\bibinfo{title}{Database of state incentives for renewables \& efficiency
  ({DSIRE})}.
\newblock \emph{\bibinfo{journal}{NC Clean Energy Technology Center}}
  (\bibinfo{year}{2016}).

\bibitem{nature_OShaughnessy}
\bibinfo{author}{O’Shaughnessy, E.}
\newblock \bibinfo{title}{Curbing rooftop solar is a poor way to promote
  equity}.
\newblock \emph{\bibinfo{journal}{Nature Energy}} \textbf{\bibinfo{volume}{7}}
  (\bibinfo{year}{2022}).

\bibitem{Willdan}
\bibinfo{title}{Cost effectiveness of {NEM} successor rate proposals under
  rulemaking 20-08-020: A comparative analysis}.
\newblock \emph{\bibinfo{journal}{California Public Utility Commission}}
  (\bibinfo{year}{2021}).
\newblock
  \urlprefix\url{https://willdan.app.box.com/s/3jpscul3lbtof5erje7f4bkqkk96uahp/file/822926041281/}.

\bibitem{nature_Borenstein}
\bibinfo{author}{Borenstein, S.}
\newblock \bibinfo{title}{It’s time for rooftop solar to compete with other
  renewables}.
\newblock \emph{\bibinfo{journal}{Nature Energy}} \textbf{\bibinfo{volume}{7}}
  (\bibinfo{year}{2022}).

\bibitem{Burcin_NEM}
\bibinfo{author}{Revesz, R.} \& \bibinfo{author}{\"{U}nel, B.}
\newblock \bibinfo{title}{Managing the future of the electricity grid:
  Distributed generation and net metering}.
\newblock \emph{\bibinfo{journal}{Harv. Env. Law Rev.}}
  \textbf{\bibinfo{volume}{41}} (\bibinfo{year}{2017}).

\bibitem{PAO_testimony}
\bibinfo{title}{Prepared testimony for a successor tariff to the current net
  energy metering tariffs}.
\newblock \emph{\bibinfo{journal}{Public Advocates Office, California Public
  Utility Commission}}  (\bibinfo{year}{2021}).
\newblock
  \urlprefix\url{https://docs.cpuc.ca.gov/PublishedDocs/SupDoc/R2008020/3899/394781553.pdf/}.

\bibitem{Sylwia}
\bibinfo{author}{Bialek, S.}, \bibinfo{author}{Dvorkin, Y.},
  \bibinfo{author}{Kim, J.} \& \bibinfo{author}{\"{U}nel, B.}
\newblock \bibinfo{title}{Who knows what: {I}nformation barriers to efficient
  {DER} roll-out}.
\newblock \emph{\bibinfo{journal}{USAEE Wor. Pap. No. 21-497}}
  (\bibinfo{year}{2021}).
\newblock \urlprefix\url{https://ssrn.com/abstract=3844269}.

\bibitem{net_billing}
\bibinfo{title}{Modernizing {California’s Net Energy Metering Program} to
  meet our clean energy goals}.
\newblock \emph{\bibinfo{journal}{California Public Utility Commission}}
  (\bibinfo{year}{2022}).
\newblock
  \urlprefix\url{https://www.cpuc.ca.gov/industries-and-topics/electrical-energy/demand-side-management/net-energy-metering/nem-revisit/net-billing-tariff-fact-sheet}.

\bibitem{FERC222}
\bibinfo{title}{{FERC} {Order No.} 2222: A new day for distributed energy
  resources}.
\newblock \emph{\bibinfo{journal}{Federal Energy Regulatory Commission
  ({FERC})}}  (\bibinfo{year}{2021}).
\newblock
  \urlprefix\url{https://docs.cpuc.ca.gov/PublishedDocs/SupDoc/R2008020/3899/394781553.pdf}.

\bibitem{NYSERDA}
\bibinfo{title}{Solar program ({NY Sun}): The value stack}.
\newblock \emph{\bibinfo{journal}{New York State Energy Research \& Development
  Authority ({NYSERDA})}}  (\bibinfo{year}{2022}).
\newblock
  \urlprefix\url{https://www.nyserda.ny.gov/All-Programs/ny-sun/contractors/value-of-distributed-energy-resources}.

\bibitem{SLMF}
\bibinfo{author}{Khan, H.}, \bibinfo{author}{Kim, J.} \&
  \bibinfo{author}{Dvorkin, Y.}
\newblock \bibinfo{title}{Risk-informed participation in {T\&D} markets}.
\newblock \emph{\bibinfo{journal}{Electric Power Systems Research}}
  \textbf{\bibinfo{volume}{202}} (\bibinfo{year}{2022}).

\bibitem{NYPSC}
\bibinfo{title}{{CASE 15-E-0302 - Proceeding on Motion of the Commission to
  Implement a Large-Scale Renewable Program and a Clean Energy Standard}}.
\newblock \emph{\bibinfo{journal}{New York Public Service Commission}}
  (\bibinfo{year}{2020}).

\bibitem{lindist}
\bibinfo{author}{{Baran}, M.} \& \bibinfo{author}{{Wu}, F.}
\newblock \bibinfo{title}{Optimal sizing of capacitors placed on a radial
  distribution system}.
\newblock \emph{\bibinfo{journal}{IEEE Tran. Pwr. Del.}}
  \textbf{\bibinfo{volume}{4}}, \bibinfo{pages}{735--743}
  (\bibinfo{year}{1989}).

\bibitem{boyd}
\bibinfo{author}{{Boyd}, S.} \& \bibinfo{author}{{Vandenberghe}, L.}
\newblock \emph{\bibinfo{title}{Convex Optimization}}
  (\bibinfo{publisher}{Cambridge Uni. Press}, \bibinfo{year}{2004}).

\bibitem{VDER_1}
\bibinfo{title}{{Case 15-E-0751 - In the matter of the value of distributed
  energy resources: Order on net energy metering transition, phase one of value
  of distributed energy resources, and related matters}}.
\newblock \emph{\bibinfo{journal}{New York Public Service Commission}}
  (\bibinfo{year}{2017}).

\bibitem{Burcin_IAEE}
\bibinfo{author}{\"{U}nel, B.}, \bibinfo{author}{Bialek, S.},
  \bibinfo{author}{Kim, J.} \& \bibinfo{author}{Dvorkin, Y.}
\newblock \bibinfo{title}{Energy transition, distributed energy resources, and
  the need for information}.
\newblock \emph{\bibinfo{journal}{IAEE Energy Forum, Third Quarter}}
  (\bibinfo{year}{2020}).

\bibitem{Robert_pricing}
\bibinfo{author}{Mieth, R.} \& \bibinfo{author}{Dvorkin, Y.}
\newblock \bibinfo{title}{Distribution electricity pricing under uncertainty}.
\newblock \emph{\bibinfo{journal}{IEEE Trans. Pwr. Syst.}}
  \textbf{\bibinfo{volume}{35}}, \bibinfo{pages}{2325--2338}
  (\bibinfo{year}{2020}).

\bibitem{conejo}
\bibinfo{author}{{Conejo}, A.} \& \bibinfo{author}{Ruiz, C.}
\newblock \bibinfo{title}{Complementarity, not optimization, is the language of
  markets}.
\newblock \emph{\bibinfo{journal}{IEEE Op. Acc. J. Pwr. Energy}}
  \textbf{\bibinfo{volume}{7}} (\bibinfo{year}{2020}).

\bibitem{MILP}
\bibinfo{author}{{Pineda}, S.} \& \bibinfo{author}{{Morales}, J.}
\newblock \bibinfo{title}{Solving linear bilevel problems using big-ms: Not all
  that glitters is gold}.
\newblock \emph{\bibinfo{journal}{IEEE Tran. Pwr. Syst.}}
  \textbf{\bibinfo{volume}{34}}, \bibinfo{pages}{2469--2471}
  (\bibinfo{year}{2019}).

\bibitem{mpec_book}
\bibinfo{author}{{Kim}, Y.}, \bibinfo{author}{{Leyffer}, S.} \&
  \bibinfo{author}{{Munson}, T.}
\newblock \bibinfo{title}{{MPEC} methods for bilevel optimization problems}.
\newblock In \emph{\bibinfo{booktitle}{Bi. opt.: Adv. \& Next Chal.}}
  (\bibinfo{year}{2019}).

\bibitem{2_kanzow}
\bibinfo{author}{{Kanzow}, C.} \& \bibinfo{author}{{Schwartz}, A.}
\newblock \bibinfo{title}{A new regularization method for {MPCCs} with strong
  convergence}.
\newblock \emph{\bibinfo{journal}{SIAM J. Opt.}} \textbf{\bibinfo{volume}{23}}
  (\bibinfo{year}{2013}).

\bibitem{relaxation_comp}
\bibinfo{author}{{Hoheisel}, T.}, \bibinfo{author}{Kanzow, C.} \&
  \bibinfo{author}{Schwartz, A.}
\newblock \bibinfo{title}{Theoretical and numerical comparison of relaxation
  methods for {MPCCs}}.
\newblock \emph{\bibinfo{journal}{Math. Prog.}} \textbf{\bibinfo{volume}{137}},
  \bibinfo{pages}{257--288} (\bibinfo{year}{2013}).

\bibitem{inexact}
\bibinfo{author}{{Kanzow}, C.} \& \bibinfo{author}{{Schwartz}, A.}
\newblock \bibinfo{title}{The price of inexactness: Convergence properties of
  relaxation methods for {MPCCs} revisited}.
\newblock \emph{\bibinfo{journal}{Math. Op. Res.}}
  \textbf{\bibinfo{volume}{40}}, \bibinfo{pages}{253--275}
  (\bibinfo{year}{2015}).

\bibitem{Scholtes}
\bibinfo{author}{{Scholtes}, S.}
\newblock \bibinfo{title}{Convergence properties of a regularization schemes
  for {MPCCs}}.
\newblock \emph{\bibinfo{journal}{SIAM J. Opt.}} \textbf{\bibinfo{volume}{11}},
  \bibinfo{pages}{918--936} (\bibinfo{year}{2001}).

\bibitem{lin}
\bibinfo{author}{{Lin}, G.} \& \bibinfo{author}{{Fukushima}, M.}
\newblock \bibinfo{title}{A modified relaxation scheme for {Mathematical
  Programs with Complementarity Constraints}}.
\newblock \emph{\bibinfo{journal}{Ann. Op. Res.}}
  \textbf{\bibinfo{volume}{133}} (\bibinfo{year}{2005}).

\bibitem{kadrani}
\bibinfo{author}{Kadrani, A.}, \bibinfo{author}{Dussault, J.-P.} \&
  \bibinfo{author}{Benchakroun, A.}
\newblock \bibinfo{title}{A new regularization scheme for {MPCCs}}.
\newblock \emph{\bibinfo{journal}{SIAM J. Opt.}} \textbf{\bibinfo{volume}{20}}
  (\bibinfo{year}{2009}).

\bibitem{ulbrich}
\bibinfo{author}{{Steffensen}, S.} \& \bibinfo{author}{Ulbrich, M.}
\newblock \bibinfo{title}{A new relaxation scheme for {Mathematical Programs
  with Equilibrium Constraints}}.
\newblock \emph{\bibinfo{journal}{SIAM J. Opt.}} \textbf{\bibinfo{volume}{20}}
  (\bibinfo{year}{2010}).

\bibitem{jean}
\bibinfo{author}{Dussault, J.-P.}, \bibinfo{author}{Haddou, M.} \&
  \bibinfo{author}{Migot, T.}
\newblock \bibinfo{title}{The new butterfly relaxation method for {Mathematical
  Programs with Complementarity Constraints}}.
\newblock In \emph{\bibinfo{booktitle}{Optimization, Variational Analysis and
  Applications. IFSOVAA 2020. Springer Proceedings in Mathematics \&
  Statistics, Springer, Singapore}}, vol. \bibinfo{volume}{355}
  (\bibinfo{year}{2021}).

\end{thebibliography}

\end{document}